\documentclass[11pt,a4paper]{article}
\usepackage{jheppub} 
\usepackage[utf8]{inputenc}

\usepackage{amsmath,amssymb,amscd,amsfonts,mathtools}
\usepackage{float} 

\usepackage{booktabs} 

\newcommand{\be}{\begin{equation}}
\newcommand{\ee}{\end{equation}}
\newcommand{\bea}{\begin{eqnarray}}
\newcommand{\eea}{\end{eqnarray}}

\renewcommand{\l}{\lambda}
\renewcommand{\t}{\tau}

\newcommand{\dd}{\mathrm{d}}

\newcommand{\mt}[1]{\textrm{\scriptsize #1}}
\def\Nc{N_\mt{c}}
\def\Nf{N_\mt{f}}
\def\Vf{V_\mt{f}}
\def\Vg{V_\mt{g}}
\def\rh{r_\mt{h}}
\def\Mp{M_\mt{p}}
\def\nB{n_\mt{b}}
\def\nq{n_\mt{q}}
\def\lah{\lambda_\mt{h}}

\def\SP{SP}
\def\AP{AP}

\title{
Quark flavors in hot and dense holographic QCD: setup and comparison to data
}

\affiliation[a]{Asia Pacific Center for Theoretical Physics,   Pohang 37673,   Korea }
\affiliation[b]{Department of Physics, Pohang University of Science and Technology,   Pohang 37673,   Korea }
\affiliation[c]{Institute for Theoretical Physics, University of Heidelberg, D-69120 Heidelberg, Germany}
\affiliation[d]{Institute of Theoretical Physics,  Chinese Academy of Sciences, Beijing 100190, China}

\author[a,b,d]{Matti J\"arvinen}
\author[c]{Toshali Mitra}
\date{July 2024}

\emailAdd{mattijarvinen@itp.ac.cn}
\emailAdd{t.mitra@thphys.uni-heidelberg.de}

\preprint{APCTP Pre2025 - 016}

\abstract{We establish a flavor dependent holographic framework for hot and dense QCD. To this end, we generalize a class of bottom-up holographic models for QCD in the Veneziano limit (V-QCD) by incorporating explicit flavor dependence. Specifically, we develop a $2+1$ flavor model characterized by two massless light quarks and a massive strange quark. Including the non-zero quark mass modifies 
the tachyon dependence of the model action, which yields a good agreement with the lattice data for thermodynamics in QCD in the low baryon number density and high temperature limit. We compare the model with various hadron resonance gas models at low temperature. We also compute the equation of state (EOS) at high density and low temperatures and find a smooth matching of this EOS with the nuclear matter EOS, suggesting significantly lower latent heat of the nuclear to quark matter transition than in earlier version of the V-QCD model. We observe that the smooth matching with the nuclear theory EOS is only applicable to a subset of potentials; furthermore, constraining the predictions of the model by limiting the potential parameters. We also identify the relevant fixed points of the model, particularly AdS$_2$ and AdS$_5$ fixed points.}

\begin{document}

\maketitle

\section{Introduction}

Exploring the properties of hot QCD matter and mapping the QCD phase diagram is a demanding task~\cite{Brambilla:2014jmp}. 
The situation is particularly challenging at nonzero baryon number density, where few experimental results are available and all known theoretical approaches face difficulties. Consequently, the locations of the phase transition lines and even relevant phases remain unclear.
However, the experimental situation is likely to significantly improve in near future thanks to heavy-ion collision experiments aimed to probe the region at nonzero density: Data from the beam energy scan at the relativistic heavy-ion collider (RHIC) is already available \cite{STAR_Bulk_BES_2017,STAR:2025zdq,STAR:2025ydk}, showing signs of a possible critical end-point of a deconfinement transition line in the phase diagram. Future experiments at facilities such as FAIR and J-PARC will push the probed region towards higher densities. 

At the same time, complementary information about cool and dense matter is obtained from observations of neutron stars. Such observations include measurements of neutron star masses and radii as well as multi-messenger observations of neutron star mergers through gravitational and electromagnetic waves. Indeed, already the gravitational wave signal from the first observed merger event GW170817~\cite{TheLIGOScientific:2017qsa,GBM:2017lvd} set highly nontrivial constraints for the equation of state (EOS) of cold QCD matter~\cite{Annala:2017llu}. Additional observations at improved sensitivity are expected to drastically improve these bounds. Therefore, combining with the neutron star observations with experimental results is likely to significantly increase our knowledge about the QCD phase diagram and the EOS.

It is important to support these experimental and observational developments
by theoretical predictions. However, while analyzing strong interactions is hard to start with, the regime of nonzero temperature and density is particularly challenging. When the temperature and baryon number chemical potential are both nonzero and near the intrinsic scale of QCD, $\Lambda_\mt{QCD}$, first-principles methods such as lattice simulations and perturbation theory are not feasible. In the absence of these methods, one can use extrapolations or interpolations~\cite{Annala:2017llu,Annala:2021gom,Altiparmak:2022bke} and models such as the Nambu-Jona-Lasinio models (see, e.g., ~\cite{Buballa:2003qv,Roessner:2006xn,Costa:2008yh}). Moreover, recently there has been promising progress with more direct approaches, such as the Dyson-Schwinger and functional renormalization group methods applied to QCD (see, e.g.,~\cite{Fu:2019hdw,Gao:2020fbl,Gunkel:2021oya}). An inherently strongly coupled alternative is to use the gauge/gravity duality (or ``holography'') to describe hot and dense QCD matter.

The gauge/gravity duality, in general, maps strongly coupled field theories to classical higher-dimensional gravitational theories. Therefore, hard problems on the field theory side may be mapped to tractable classical calculations on the gravity side. However, in the case of QCD such a duality has not been established and currently there are no promising attempts to derive this kind of duality in any precise setting. It is not even known if such a duality exists in some well-defined sense. Nevertheless, gauge/gravity duality has proven to be a useful tool to analyze observables for which direct QCD predictions are hard to obtain, such as viscosities of hot QCD plasma~\cite{Policastro:2001yc,Kovtun:2004de}. Holographic models are divided roughly into two classes: the top-down and the bottom-up models. The top-down models, such as the Witten-Sakai-Sugimoto model~\cite{Witten:1998zw,Sakai:2004cn,Sakai:2005yt}, are based on concrete fixed string theory backgrounds (typically in ten-dimensional gravity setups) with certain brane configurations. Such models are dual to a fixed field theory that is reminiscent of QCD but differs from it, e.g., by having additional states. The bottom-up models are generic models constrained by symmetry and general principles of holography. They typically contain a high number of parameters that can be adjusted to mimic the properties of QCD. However, there is no control on what precisely this adjusting procedure means on the QCD side, and one cannot specify which field theory exactly is being modeled even if it appears to be similar to QCD. Nevertheless, such models have been shown to give, for example, a surprisingly good description of meson spectra in QCD~\cite{Erlich:2005qh,DaRold:2005mxj,Karch:2006pv} and simultaneous precise model for the thermodynamics and glueball spectrum of the Yang-Mills theory~\cite{Gursoy:2009jd,Panero:2009tv}.

The goal of this article is to further develop a data-driven bottom-up approach for QCD, focusing on physics at nonzero temperature and/or baryon number density. We employ a specific holographic setting, the V-QCD model~\cite{Jarvinen:2011qe,Jarvinen:2021jbd}, which is a rich bottom-up model for QCD, including separate sectors for gluon and (fully backreacted) quarks. The dynamics of the model is controlled by the choice of several potentials, i.e., functions of a scalar field, which are adjusted to precisely mimic the thermodynamics and other properties of QCD.

Specifically, in this article, we focus on the flavor dependence of the V-QCD model. That is, we develop an extension of the model which allows one to describe deconfined quark matter with nonzero quark masses and chemical potentials independently for each quark flavor. A simple version of the flavor-dependent V-QCD model was already used to analyze the bulk viscosity of dense QCD matter in~\cite{CruzRojas:2024etx}. 
Here we develop this model further. We present a general setup allowing a dependence on the masses and chemical potentials of an arbitrary number of quarks, but focus on the case of $\Nf = 2+1$ flavors, i.e., two light quarks and the strange quark, carrying out a detailed comparison to lattice results for the EOS of QCD. While the masses of the light quarks (up and down flavors) may affect significantly some specific features such as the order of the deconfinement phase transition (see, e.g.,~\cite{Guenther:2020jwe}), in general their effect of the EOS is expected to be small as their values are tiny compared to $\Lambda_\mt{QCD}$ and the crossover temperature in QCD. In this work, we will simply set the light quark masses to zero. However, the mass of the strange quark is comparable to the critical temperature, and its effect on the EOS is significant, as seen for example in the flavor dependence of the lattice data for quark number susceptibilities~\cite{Borsanyi:2011sw,HotQCD:2012fhj}. 
Some other observables, such as the bulk viscosity due to weak interactions coupled to QCD matter analyzed in~\cite{CruzRojas:2024etx}, can be sensitive to the strange quark mass, so a flavored model is required to obtain reliable predictions. Moreover, various holographic models for QCD have a modulated instability near the region where the critical point of the deconfinement transition of QCD is expected to lie~\cite{CruzRojas:2024igr,Demircik:2024aig}, which also appears to be sensitive to the strange quark mass. The same applies to the location of the critical point in the models in the same class (see, e.g.,~\cite{DeWolfe:2010he,Knaute:2017opk,Critelli:2017oub,Demircik:2021zll,Cai:2022omk,Ecker:2025vnb}).

Implementing full flavor dependence gives rise to new observables even when the quark masses are set to zero, such as nondiagonal susceptibilities $\chi_{ij} = \frac{\partial^2 p}{\partial\mu_i\partial\mu_j}$, where $\mu_i$ are the chemical potentials for different flavors, and in general the EOS away from flavor symmetry, including isospin imbalance and symmetry energy. 
Moreover, implementing quark flavors separately will allow one to turn on a magnetic field that couples correctly to the quarks, i.e., according to their electric charges rather than just uniformly to all quarks.
There are several earlier works discussing some aspects of flavor dependence both in top-down~\cite{Hata:2007tn,Erdmenger:2007vj,Hashimoto:2009st,Kim:2011gw,Liu:2016iqo,Fujii:2020jre} and bottom-up~\cite{Shock:2006qy,Afonin:2013npa,Li:2016smq,Ballon-Bayona:2017bwk,Fang:2018vkp,Chen:2018msc,Chen:2021wzj,Ahmed:2024rbj} holography, 
but these models have not been adjusted to reproduce the detailed flavor dependence of the QCD EOS revealed by lattice analysis. The main goal of this article is to fix this situation. 

In order to establish the $\Nf = 2+1$ model for QCD, we need to carry out a careful study of the background solutions in the gravity side of the duality, which correspond to various phases on the field theory side. We begin by identifying the simplest solutions following~\cite{Alho:2013hsa}: the fixed point solutions for which the scalar fields in the gravity are set to constants. As it turns out, we find two classes of solutions, which we call the AdS$_5$ and AdS$_2$ points. These fixed point solutions then serve as the end points of a more general class of solutions, for which the scalar fields flow in the holographic direction, implementing the renormalization group flow expected in QCD. We show how the phase diagram of the model arises from the competition between different classes of flowing geometries, but also remark how the fixed point solutions are relevant for the behavior of the flowing solutions at asymptotically high and low temperatures.

The article is organized as follows. In Sec.~\ref{sec:setup} we define the flavored V-QCD setup, including the dictionary, details of the action, Ansatz for the background geometries, equations of motion and general formulas for thermodynamics. In Sec.~\ref{sec:const_scalars} we identify the possible constant scalar fixed point solutions of the flavored setup, which appear as endpoints of flows in more complex solutions determining the phase diagram, and also discuss tachyonic perturbations around the fixed points. In Sec.~\ref{sec:zeromu} we discuss how the EOS is computed from the model at low density in the deconfined phase, and analyze the role of the fixed point solutions in the final geometries. Furthermore, we fix model parameters by comparing to lattice data, and also compare the results to hadron resonance gas (HRG) models. In Sec.~\ref{sec:finitemu} we discuss model predictions at nonzero density, focusing on the quark matter at low temperatures, and also compare our EOS to nuclear theory models and model-independent parametrizations. Finally, we summarize our results and discuss future directions in Sec.~\ref{sec:conclusion}.

\section{Setup}\label{sec:setup}

We discuss here a generalized setup for the V-QCD model which includes flavors and the possibility to turn on quark masses independently for each flavor. The original setup for quark degrees of freedom~\cite{Jarvinen:2011qe,Alho:2013hsa}, which is based on earlier work~\cite{Bigazzi:2005md,Casero:2007ae}, in principle already includes the flavor structure, but the details on turning on the masses were only discussed recently in the context of analyzing bulk viscosity at nonzero density~\cite{CruzRojas:2024etx}. In this article, in order to precisely fit the available data for QCD thermodynamics on the lattice, we present a further generalization of this setup.

The V-QCD model is obtained, in principle, by considering QCD with $\Nf$ flavors and SU$(\Nc)$ gauge group in the Veneziano limit~\cite{Veneziano:1976wm}, where $\Nc$ and $\Nf$ are taken to infinity keeping their ratio $x\equiv \Nf/\Nc$ fixed, {therefore ensuring that the quarks fully backreact to the gluon dynamics}. There is an inspiration from five-dimensional noncritical string theory making use of this limit~\cite{Gursoy:2007cb,Jarvinen:2011qe}. However, no precise derivation of the model from string theory exists, so in the end one needs to switch to bottom-up approach where one generalizes the action by introducing extra parameters and potentials which are then determined by comparing to QCD results. In particular, rather than considering the Veneziano limit, one in practice simply inserts $\Nc=3$ and $\Nf=2$ or $\Nf=3$ when comparing to QCD.

In this article, we want to do a more detailed analysis of flavors in the holographic model. If we want to follow the stringy picture with the Veneziano limit, the number of flavors is large in the model. In order to compare to lattice data, which typically has $\Nf=2+1$, i.e., two light quarks and the strange quark, one can divide this large number of flavors into three groups corresponding to the physical quark flavors. However, we choose to simply set $\Nf$ and $\Nc$ to three from the start and ignore corrections in $1/\Nc$ or $1/\Nf$. {We do not see an easy way to control these corrections. However, lattice studies in pure Yang-Mills theory \cite{Panero:2009tv} indicate that corrections in $1/N_c$ are small. For full QCD similar analysis does not exist, but a natural expectation is that the corrections remain to be small also in the Veneziano limit.}

To be precise, we will consider {also in holography} the case of $\Nf=2+1$ which means that two of the quarks (the light quarks) will be massless whereas the third (the strange quark) has a nonzero mass. {Note that since we also set $\Nc=3$, backreaction of the quarks to the gluons remains significant.}

\subsection{Dictionary and action}

Let us then review the dictionary of the model. The relevant five-dimensional fields are those dual to the most important (relevant and marginal) operators in QCD:
\begin{itemize}
 \item The metric $g_{MN}$, which is dual to the energy momentum tensor $T_{\mu\nu}$ of QCD.\footnote{Our notation is such that five-dimensional (four-dimensional) Lorentz indices are denoted by uppercase Latin (lowercase Greek) letters, respectively. It is easiest to set up the dictionary in a gauge where $g_{rM} = 0$, where $r$ stands for the holographic direction, and also $A_{L/R\, r}^{ij} =0$. This eliminates the extra components of the metric and gauge fields appearing in the gravity side with respect to the field theory duals.} The source is therefore the metric of the field theory (which will be set here to the Minkowski metric).
 \item The dilaton $\phi$, which is dual to the gluon field strength tensor squared $\mathrm{tr}G_{\mu\nu}G^{\mu\nu}$, where the trace is over the color indices. The source for this operator, and therefore the source of the field in gravity, is the gauge coupling $g$ of QCD. In holographic context, it is typically most convenient to use the 't Hooft coupling $\lambda_t = g^2\Nc$.
 \item The tachyon, $T_{ij}$, which is a complex matrix in flavor space ($i,j=1,\ldots , \Nf$). It is dual to the quark bilinear $\bar \psi^j \psi^i$ so that the source is the complex quark matrix, $M_{ij}$. 
 \item The left and right handed gauge fields $A^{ij}_{R/L\,M}$, reflecting the fact that the chiral symmetry U$(\Nf)_L\times$U$(\Nf)_R$ is promoted to a gauge symmetry on the gravity side of the correspondence. In this article, it is enough to consider the vectorial field $A_M^{ij}$, which is dual to the current $\bar \psi^i \gamma_\mu \psi^j$. The sources for these currents are external fields and, importantly, the chemical potentials for different quark flavors. 
\end{itemize}

The V-QCD model is obtained by combining a gluon sector, described on the gravity side by the Improved Holographic QCD (IHQCD) model~\cite{Gursoy:2007cb,Gursoy:2007er,Gursoy:2010fj}, and a quark sector arising from a  tachyonic D-brane picture~\cite{Bigazzi:2005md,Casero:2007ae}. Therefore the action of the model contains two terms,
\begin{equation}
    S_\mt{V-QCD} = S_\mt{g} + S_\mt{f} \ ,
\end{equation}
reflecting the two sectors of the model. The first term is five-dimensional Einstein-dilaton gravity,
\begin{equation}
    S_\mt{g} = \Mp^3 \Nc^2 \int \dd^{5} x\sqrt{-g}\left[R-\frac{4}{3} g^{M N} \partial_{M} \phi \partial_{N} \phi+\Vg(\phi)\right] \ ,
\end{equation}
which (with a proper choice~\cite{Gursoy:2009jd} for the dilaton potential $\Vg$ and the Planck mass $\Mp$) implements the IHQCD model. Note that this action governs the dynamics of the metric and the dilaton, which are indeed the duals of the gluonic operators in the dictionary. 

The flavor sector~\cite{Bigazzi:2005md,Casero:2007ae} is inspired by the action of a pair of space filling tachyonic $D4$--$\overline{D4}$ branes. It contains both Dirac--Born--Infeld (DBI) and Chern--Simons (CS) terms. The CS terms are important for the description of anomalies~\cite{Arean:2016hcs,Jarvinen:2022mys}, baryons~\cite{Jarvinen:2022mys,Jarvinen:2022gcc} and modulated instabilities~\cite{Demircik:2024aig,Demircik:2021zll,CruzRojas:2025fzs} in the model, but irrelevant for the current article. Therefore, we only include the DBI term, which reads
\begin{align} \label{eq:Sfgen}
S_\mt{f}&=-\Mp^3 \Nc \int \dd^5x\,\mathrm{Tr}\left[\Vf(\phi,T)\sqrt{-\det (g_{MN} + w(\phi,T) F_{MN}+\kappa(\phi)\partial_M T\partial_N T)}\right] \ . 
\end{align}
Here the trace is over the flavor indices, $F_{MN}$ is the field strength tensor of the vectorial gauge field $A_M$, and three additional potential functions $\Vf(\phi,T)$, $w(\phi,T)$, and $\kappa(\phi)$ were introduced. Because we only turned on the vectorial gauge fields, the DBI actions of the $D4$ and the $\overline{D4}$ branes merge into a single DBI action. Note that, unlike~\cite{CruzRojas:2024etx}, we consider coupling functions $w$ for the gauge which depend on the tachyon field $T$. Moreover, in order to properly define the action~\eqref{eq:Sfgen} one should specify how the trace over the flavors is defined which is a nontrivial task because the different flavor fields do not necessarily commute. However in this article we will only consider backgrounds where the fields commute and the trace is therefore unambiguous. Note also that as in earlier literature, the flavor action only contains a single flavor-trace. Addition of multi-trace contributions is discussed in the conclusions.

\subsection{Background solutions}

In order to proceed, we specify our Ansatz for all the five-dimensional fields appearing in the action. For the purposes of this article, it is enough to study homogeneous and time-independent backgrounds. First, for the metric we write
\begin{equation} \label{metric}
\mathrm{d} s^{2}=e^{2A(r)}\left[-f(r) \mathrm{d} t^{2}+\mathrm{d} \mathbf{x}^{2}+\frac{\mathrm{d} r^{2}}{f(r)}\right]\ , 
\end{equation}
where $A(r)$ is the scale factor, corresponding to the logarithm of the energy scale in field theory~\cite{Gursoy:2007cb,Gursoy:2007er}, and $f(r)$ is the blackening factor.
In these coordinates, the boundary, where one makes connection to the field theory, is located at a finite value of $r$ which we set to be zero. We will set up the action such that near the boundary the metric is asymptotically AdS$_5$ and the blackening factor approaches one at the boundary, $f(0)=1$.

We only consider homogeneous backgrounds here, so that all fields are functions of the holographic coordinate $r$ only. We replace the dilaton field by its exponential
\begin{equation}
   \lambda(r) \equiv \exp(\phi(r)) \ ,
\end{equation}
because it is this field that is identified as the 't Hooft coupling near the boundary~\cite{Gursoy:2007cb}.
We take the flavor-dependent fields to be diagonal in flavor indices: $T^{ij}=\tau_i\delta^{ij}$ and $(A_M)^{ij} = A^{(i)}_M \delta^{ij}$. Therefore the flavor action simplifies \footnote{{This simplified action excludes non-diagonal (e.g. isospin) chemical potentials and also other sources which are off-diagonal in flavor, including phases induced by non-Abelian external fields. In principle also spontaneously formed non-diagonal states such as p-wave condensates~\cite{Gubser:2008wv,Jarvinen:2024wsn} are excluded, but these states typically require non-diagonal chemical potentials.}} 
to
\begin{align} \label{eq:Sfflavors}
S_\mt{f} & = -\Mp^3 \Nc \sum_{i=1}^{\Nf} \int \dd^5x\,\Vf(\lambda,\tau_i)\sqrt{-\det (g_{MN}  + w(\lambda,\tau_i) F_{MN}^{(i)}+\kappa(\lambda)\partial_M \tau_i\partial_N \tau_i)} \ ,
\end{align}
where $F^{(i)}_{MN}$ is the field strength tensor for the field $A^{(i)}_M$. Furthermore, we use the radial gauge, $A_r^{(i)}=0$. 

At finite temperature, there are two classes of geometries that we need to consider:
\begin{enumerate}
    \item ``Thermal gas'' geometries with $f(r)=1$ and no horizon. As we discuss in Sec.~\ref{sec:bgs_zero_density}, there are several types of geometries in this class. However, the ones which turn out to be relevant for the final phase diagram are those for which the geometry ends in a ``good'' kind of singularity in the classification of~\cite{Gubser:2000nd}, see~\cite{Gursoy:2007cb,Gursoy:2007er,Jarvinen:2011qe}. These geometries are duals to confined phases in field theory.
    \item Black hole backgrounds where $f(r)$ depends on the coordinate and there is a horizon, $f(\rh)=0$ at some finite value $r=\rh$. Black holes are dual to deconfined phases.
\end{enumerate}
In this article, we will be focusing on the black hole geometries. In case any of the quark masses is set to zero, such that the corresponding tachyon component does not have a source, both the confined and deconfined geometries have variations where the tachyon component is either zero or nonzero (indicating spontaneous breaking of some of the chiral symmetry)~\cite{Alho:2012mh,Alho:2013hsa}. Examples of such solutions will be discussed below.

The thermal gas solutions have trivial temperature dependence (e.g., the pressure is temperature independent and the entropy vanishes) and therefore also a smooth zero temperature limit. However, the black hole solutions do depend on temperature. In the zero temperature limit, the IR geometry becomes asymptotically AdS$_5$ or AdS$_2 \times \mathbb{R}^3$ as discussed in the case of V-QCD in~\cite{Alho:2013hsa,Hoyos:2021njg}. We will discuss how this IR analysis is generalized to the flavor dependent case (with differences between flavors arising from quark masses) below.

In order to pin down the model, the potentials $\Vg$, $\Vf$, $\kappa$, and $w$ in the above actions need to be specified. All these potentials have already been determined for the unflavored model, so in an optimal scenario, we could simply use the same potentials in the flavored setup. This was actually done in~\cite{CruzRojas:2024etx}, where we estimated the bulk viscosity by using the flavored V-QCD setup. However, this approach does not work perfectly, because the model obtained this way is in clear disagreement with the flavor dependence of quark number susceptibilities seen in lattice analysis. (We will discuss this dependence in more detail below.) Therefore, the functions need to be adjusted slightly. 

To be more precise, the comparison of the potentials with QCD data has several steps. As it turns out, qualitative agreement with features of QCD strongly constrains the asymptotic behavior of the potentials both at weak and strong coupling $\lambda$. Having AdS$_5$ asymptotics near the boundary with similar renormalization group flow as in QCD requires that the potentials approach constants at $\lambda=0$, with power corrections in $\lambda$~\cite{Gursoy:2007cb,Gursoy:2007er,Jarvinen:2011qe}. Conversely, requiring agreement with low-energy features, e.g., confinement, discrete spectra, linear trajectories of radial meson excitations,
and qualitatively correct phase diagram, fixes the asymptotics of the potentials at large $\lambda$~\cite{Arean:2013tja,Jarvinen:2015ofa,Jarvinen:2021jbd}. Working potentials have typically power asymptotics in $\lambda$ with multiplicative logarithmic corrections. All these qualitative requirements are unaffected by the inclusion of flavor dependence. Therefore, we use even the same Ans\"atze as in the unflavored case, apart from small adjustments of the dependence on the tachyon field $\tau_i$. These adjustments and small changes in the parameters of the potentials of the DBI action is enough to fit well the flavor dependence seen on the lattice.

There are already different kinds of unflavored fits available in the literature, which fit the lattice data for QCD thermodynamics~\cite{Jokela:2018ers,Ishii:2019gta}, meson or glueball spectra and decay constants~\cite{Amorim:2021gat}, or both~\cite{Gursoy:2009jd,Jarvinen:2022gcc}. Since the main goal of the current work is to construct a model for the thermodynamics and the EOS of QCD, we choose to use as a starting point the fits of~\cite{Jokela:2018ers} which give the best description of lattice data at nonzero temperature.  
Apart from the parameters of the potentials, also the Planck mass $\Mp$ and the overall energy scale $\Lambda$, which is defined though the near-boundary expansions of the geometry in Appendix~\ref{app:asymptotics}, are fitted to data. The details of fit will be discussed in Sec.~\ref{sec:lattice}, and the fit results can be found in Appendix~\ref{app:st_pots}. In addition, following the holographic dictionary, the quark masses are defined through the boundary expansion of the tachyon components $\tau_i$, see Appendix~\ref{app:asymptotics}.

\subsection{Equations of motion and thermodynamics} \label{sec:thermo}

We then discuss the equations of motion arising from the action and how they are used to compute thermodynamic quantities. As the thermodynamics of the thermal gas solutions is simple, we write down the formulas for black hole geometries, which is the main focus of this article.

The finite density setup is obtained by turning on temporal components $\Phi_i \equiv A_t^{(i)}$ of the gauge fields.  As the component for each quark flavor is independent, we can describe a configuration where different flavors have different densities $n_i$. We denote the quark and baryon number densities as 
\begin{equation}\label{eq:nqdef}
    \nq = \sum_{i=1}^{\Nf} n_i = \Mp^3 \Nc \sum_{i=1}^{\Nf} \bar n_i \ , \qquad
    \nB = \frac{\nq}{\Nc} = \Mp^3 \sum_{i=1}^{\Nf} \bar n_i \ ,
\end{equation}
where by the equation of motion of the gauge field~\cite{Alho:2013hsa}
\begin{equation}
    \bar n_i \equiv \frac{n_i}{\Mp^3\Nc} = - \frac{ e^{A}   \Vf(\lambda,\tau_i)w(\lambda, \tau_i)^2\dot \Phi_i}{\sqrt{ 1+e^{-2A}f \kappa (\lambda) \dot \tau_i^2-e^{-4 A}w(\lambda, \tau_i)^2\dot \Phi_i^2 }}
\end{equation}
is independent of $r$.
Here dots denote derivatives with respect to $r$. This equation can be solved for the derivative,
\begin{align} \label{eq:Phidoteq}
    \dot{\Phi}_i = -\frac{\bar n_i e^{-A}}{ \Vf(\lambda,\tau_i) w(\lambda,\tau_i)^2} \frac{\sqrt{1+e^{-2A} f \kappa \dot{\tau}_i^2}}{\sqrt{1+K_i}} 
\end{align}
where 
\begin{equation}
K(\lambda, \tau_i)  \equiv K_i  = \frac{ e^{-6 A}  \bar  n_i^2}{ w(\lambda,\tau_i)^2 \Vf(\lambda,\tau_i)^2} 
\end{equation}
and $i$ stands for different flavor dependence. The {temporal components of the gauge} fields $\Phi_i$ are required to vanish at the horizon $r=\rh$ {to ensure regularity}. Therefore integrating~\eqref{eq:Phidoteq} we find the values of the sources, i.e., the chemical potentials:
\begin{equation}
    \mu_i = \left.\Phi_i\right|_{r=0} =  \bar n_i\int_0^{\rh} \dd r \frac{e^{-A}\sqrt{1+e^{-2A}f \kappa(\lambda)\dot \tau_i^2}}{\Vf(\lambda,\tau_i)w(\lambda, \tau_i)^2\sqrt{1+K_i}} \ .
\end{equation}
From here we also obtain a simple expression for the susceptibilities at zero density~\cite{Alho:2013hsa},
\begin{equation}
    \chi^{i} \equiv \frac{\partial^2 p}{\partial \mu_i^2} = \frac{\partial n_i}{\partial \mu_i} = M_p^3 \Nc \left[\int_0^{\rh} \dd r \frac{e^{-A}\sqrt{1+e^{-2A}f \kappa(\lambda)\dot \tau_i^2}}{\Vf(\lambda,\tau_i)w(\lambda, \tau_i)^2}\right]^{-1} \ .
\end{equation}
There is no summation over $i$ here, and partial derivatives are taken keeping the temperature and other chemical potentials fixed. These susceptibilities are the diagonal components of the flavored susceptibility matrix $\frac{\partial^2p}{\partial\mu_i\partial\mu_j}$, where derivatives are taken keeping the temperature and other quark chemical potentials fixed. The nondiagonal components vanish at zero density. However, they are nonzero at finite density~\cite{CruzRojas:2024etx}.

The above formulas are completed by the usual black hole thermodynamics: the temperature and entropy are given by the surface gravity and the area of the black hole,
\begin{equation} \label{eq:BHthermo}
    T = \frac{1}{4\pi}\left|\dot{f}(\rh)\right|\ , \qquad s = 4\pi \Mp^3 \Nc^2 e^{3 A(\rh)} \ , 
\end{equation}
respectively.

The equations of motion for the other fields are given in Appendix~\ref{app:EOM}, where we eliminated the gauge field by using~\eqref{eq:Phidoteq}. Actually, note that this elimination can be made explicit at the action level by carrying out a Legendre transformation as follows. We first write the action~\eqref{eq:Sfflavors} for our Ansatz,
\be \label{eq:Sfsubst}
S_\mt{f}=-\Mp^3 \Nc V_4 \int_0^{\rh} \dd r\,\sum_{i=1}^{\Nf}e^{5A}\Vf(\phi,\tau_i)\sqrt{ 1+e^{-2A}f \kappa (\lambda) \dot \tau_i^2-e^{-4 A}w(\lambda, \tau_i)^2\dot \Phi_i^2 } \ ,
\ee
where $V_4$ is the volume of space-time. The Legendre transformed action is defined as
\be
 \widetilde{S}_\mt{f} \equiv -V_4 \sum_{i=1}^{\Nf}\mu_i n_i + S_\mt{f} \ ,
\ee
where the first term may be written as an integral
\be
  -\sum_{i=1}^{\Nf}\mu_i n_i = \int_{0}^{\rh} \sum_{i=1}^{\Nf}\dot \Phi_i n_i 
\ee
where we used the fact that $\Phi_i$ vanish at the horizon. Using~\eqref{eq:Phidoteq} to eliminate $\dot \Phi_i$ gives the final expression
\be \label{eq:Legendre_final}
 \widetilde{S}_\mt{f} = -\Mp^3 \Nc V_4 \int_0^{\rh} \dd r\,\sum_{i=1}^{\Nf}e^{5A}\Vf(\phi,\tau_i)\sqrt{1+K_i}\sqrt{ 1+e^{-2A}f \kappa (\lambda) \dot \tau_i^2} \ .
\ee
As one can check, the equations of motion given in Appendix~\ref{app:EOM}, where the gauge fields have been eliminated in favor of the charges, directly follow from this flavor action.\footnote{To be precise, this would require substituting the gauge-fixed metric~\eqref{metric} also to the gravity action before deriving the equations of motion. The constraint equation~\eqref{eq:constraint} cannot be obtained this way. However this is irrelevant to our point here, i.e., how the structure of the flavor terms arises.}

\section{Constant scalar solutions and fixed points} \label{sec:const_scalars}

Here, we discuss the fixed-point solutions for the modified V-QCD model. They give end points to more general solutions, where the dilaton field $\lambda$ flows, which models the renormalization group flow of the coupling in QCD. The flow solutions will be reviewed in Sec.~\ref{sec:zeromu} and~\ref{sec:finitemu}.

We start by discussing the general case with an arbitrary number of flavors, but then focus on fixed point relevant for the $\Nf=2+1$ case where the strange quark mass is non-zero.
Generally, fixed points are the scaling solutions associated with either fixed scalars or scalars diverging to infinity, which are often hyperscaling-violating solutions.
Such fixed point solutions have been studied for the flavor-independent V-QCD setup in~\cite{Alho:2013hsa}.
These solutions can be classified into two classes depending on the nature of the warp function $A(r)$. For a non-trivial $A(r)$, the fixed point corresponds to AdS$_5$ type geometry, representing (for example) the UV fixed point in the dual theory. Conversely, for a constant $A(r)$, we obtain the AdS$_2$ type geometry, which indicates emergent scaling behaviour in the IR~\cite{Faulkner:2009wj}. 
Here we generalize the flavor-independent discussion of~\cite{Alho:2013hsa} to the flavor-dependent case and make additional remarks. 

To determine the fixed points, we set the scalars $(\lambda,\tau_i)$ to constants $(\lambda_0, \tau_{i0})$. This is with the understanding that the values may also be infinite, i.e., considering the limit of either $\lambda_0 \to \infty$ or $\tau_{i0} \to \infty$.
These replacements  simplify the equation of motion (appendix \ref{app:EOM}) in terms of an effective potential $V_\mt{eff}$ defined as~\cite{Hoyos:2021njg}, 
\begin{align}
    V_\mt{eff}\left(\lambda_0, \{\tau_{k0}\},\{\bar n_k\},A\right) \equiv V_\mt{eff0} = \Vg(\lambda_0) - \frac{1}{\Nc}\sum_{i=1}^{\Nf} \Vf(\lambda_0,\tau_{i0}) \sqrt{1+ K(\lambda_0, \tau_{i0},\bar n_i,A)}
\end{align}
where $x=\Nf/\Nc$ and
\be
K(\lambda_0, \tau_{i0},\bar n_i,A) \equiv K_{i0} =  \frac{\left( e^{-3 A}  \bar  n_i \right)^2}{\left( w(\lambda_0,\tau_{i0})\Vf(\lambda_0,\tau_{i0})\right)^{2}} \ .
\ee 
Note that the form of the effective potential follows from the Legendre transformed action in~\eqref{eq:Legendre_final}.

As the scalars are set to be constants,  the simplified equations of motion read as
\begin{align} \label{Eq:f}
        \ddot{f} + 3 \dot{A} \dot{f} - \frac{e^{2 A}}{3} \partial_A V_\mt{eff0}  = 0
\end{align}
\begin{align} \label{Eq:lpp}
    \partial_{\lambda_0} V_\mt{eff0}  = 0 
    \end{align}
    \begin{align} \label{Eq:t}
    \partial_{\tau_{i0}} V_\mt{eff0}= 0
\end{align}
\begin{align} \label{Eq:lp}
   12 \dot{A}^2 + 3 \frac{\dot{A} \dot{f}}{f} - \frac{e^{2 A}}{f} V_\mt{eff0} = 0 
   \end{align}
\begin{align} \label{Eq:A}
      \ddot{A}  - \dot{A}^2 = 0 
    \end{align}
The independent solutions of the last equation correspond to the AdS$_5$ result $ A = - \log(r)+\mathrm{const}$ and  AdS$_2$ condition $A = \mathrm{const}$. Note that, up to the symmetries of the solution (see Appendix~\ref{app:EOM}), these expressions give all the solutions to this equation.  

\subsection{AdS$_2$ fixed points} \label{sec:fixedpoints}

For AdS$_2$ type solutions, the warp factor $A(r)$ is a constant, $A(r) = A_0$. This implies that the factors $K_{i0}$ are also constants.
Using~\eqref{eq:Phidoteq}, the solution of the gauge field becomes~\cite{Alho:2013hsa}, 
\begin{align}
   \Phi_i(r) = \mathcal{E}_{i0} r + \mu_{i0}\ , \qquad    \mathcal{E}_{i0} = -\frac{\bar n_i e^{-A_0}}{V_{\mt{f}}(\lambda_0,\tau_{i0}) w(\lambda_0,\tau_{i0})^2} \frac{1}{\sqrt{1+K_{i0}}}  \ .
\end{align}
The general solution of the blackening factor is given by,
\begin{align}
    f(r) = c_2 + c_1 r + r^2  \frac{e^{2 A_0}}{6} \partial_{A_0} V_\mt{eff0} \ . 
\end{align}
We require that $\Vf$ is positive, which leads to $\partial_{A_0} V_\mt{eff0}>0$. Depending on the values of the coefficients $c_i$, the blackening factor may or may not have roots. We are only interested in the case where it has roots. By translation symmetry, we may set one of the roots to be at $r=0$, meaning that $c_2=0$. Then regularity of the gauge field solution sets also $\mu_{i0}=0$. The remaining parameter $c_1$ is related to the temperature. If $c_1<0$, the physical region (with $f(r)>0$) is at $r<0$, and the temperature is $T=-c_1/(4\pi)$. Actually, the AdS$_2$ solution is the one with $c_1=0$ (in addition to $c_2=0$). As we discuss in Sec.~\ref{sec:finitemu}, this geometry appears as the end point of vacuum solutions with flowing dilaton $\lambda$ at nonzero density. The AdS$_2$ solutions with nonzero $c_1$ are also relevant for flowing solutions as they represent the IR geometry for black hole solutions at small temperatures (and nonzero density), $T/\Lambda \ll 1$. At higher temperatures, i.e., for generic values of $c_1$, the flow of the coupling breaks the correspondence between the flowing finite temperature solutions and AdS$_2$ black hole solutions.

So far we only used the equations~\eqref{Eq:f} and~\eqref{Eq:A}.
The other three equations are constraint equations that determine when the solutions can be realised,
\begin{align}
\label{eq:Vconstr}
    & V_\mt{eff0} = 0 \\
\label{eq:laconstr}
   & \partial_{\lambda_0} V_\mt{eff0} = 0  \\
   & \partial_{\tau_{i0}} V_\mt{eff0}= 0 \label{eq:tauconstr}
\end{align}
Here the last constraint~\eqref{eq:tauconstr} is expected to fix the values of the tachyon components $\tau_{i0}$, while the other two set constraints for $\l_0$ and the charges $\bar n_i$. They essentially only fix $\l_0$ and the total charge, leaving many free parameters. In this aspect the flavor dependent fixed point is rather different from the unflavored fixed point considered in~\cite{Alho:2013hsa,Hoyos:2021njg}, where all parameters were uniquely fixed. This is natural because (as we discuss in Sec.~\ref{sec:finitemu}) the AdS$_2$ solutions give the IR endpoints of the flowing geometries at zero temperature and nonzero density. In the setup with flavor dependence, such geometries naturally depend on the individual densities of the quark flavors, giving rise to a larger parameter space.

Let us also point out that it is natural to discuss the fixed points in terms of the dimensionless ratios
\begin{equation}
 \tilde n = \frac{4\pi \nq}{s} =  \frac{e^{-3 A_\mt{h}}}{\Nc} \sum_{i=1}^{\Nf}\bar n_i \ , \qquad 
 \tilde n_i = e^{-3 A_\mt{h}}\bar n_i \ ,
\end{equation}
so that relation between $\tilde n$ and $\tilde n_i$ is 
\begin{equation}
 \tilde n = \frac{1}{\Nc}\sum_{i=1}^{\Nf}\tilde n_i \ .
\end{equation}
Below we will focus on setups with $\Nf=\Nc$ so that the right-hand side of this relation simply boils down to the average of the $\tilde n_i$'s. For the AdS$_2$ fixed points, as $A$ is constant, we find that
\begin{equation}
 K_{i0} = \frac{\tilde n_i^2}{ w(\lambda_0,\tau_{i0})^2\Vf(\lambda_0,\tau_{i0})^2} \ .
\end{equation}
Consequently, the dependence of the effective potential on $A_0$ is absent when the potential is expressed in term of $\tilde n_i$ rather than $\bar n_i$. Therefore the conditions~\eqref{eq:Vconstr} and~\eqref{eq:laconstr} constrain directly the values of $\tilde n_i$, while the values of $\bar n_i$ additionally depend on the scale $A_0$.

\subsection{AdS$_5$ fixed points}

For AdS$_5$ type solutions, the warp factor $A(r)$ is given by 
\begin{align}
    A(r) = \log \frac{\ell}{r} \ ,
\end{align}
where $\ell$ is the AdS radius. Note that for such solutions, in general, the factor $K_{i0}$ depends on $r$:
\be
 K_{i0}= \frac{ r^6  \bar n_i^2}{ \ell^6w(\lambda_0,\tau_{i0})^2\Vf(\lambda_0,\tau_{i0})^{2}} 
\ee
implying that the effective potential depends on the coordinate in a complicated manner. This prevents fixed point solutions in the general case (for example, as the condition~\eqref{Eq:lpp} cannot be satisfied). However, special fixed point solutions are possible, where the $r$-dependence is absent. This can happen if\footnote{There is also the possibility $\Vf(\l,\t_{i0}) \to 0$ which indeed leads to simplified equations of motion, but as it turns out, is not enough to allow fixed point solutions.} $\bar n_i=0$ or $w(\l_0,\t_{i0}) \to \infty$ at the fixed point for all flavors $i$. The latter condition will not be possible for potentials in our model, so we are only left with the full neutral case, i.e., $\bar n_i =0$ for all $i=1, \ldots \Nf$. Then the equations~\eqref{Eq:f} and \eqref{Eq:lp} are satisfied for the ``standard'' AdS solution, i.e.,
\begin{align} \label{eq:fAdS5}
   f(r) = 1 - \left(\frac{r}{\rh}\right)^4
\end{align}
if the AdS radius is given by
\be \label{eq:ellAdS5}
 \ell = \sqrt{\frac{12}{V_\mt{eff0}}}\ ,
\ee
where $V_\mt{eff0}$ is now simply a constant.

In this case the remaining constraints are given by
\begin{align}
\label{eq:laconstr2}
   & \partial_{\lambda_0} V_\mt{eff0} = 0  \\
   & \partial_{\tau_{i0}} V_\mt{eff0}= 0 \label{eq:tauconstr2}
\end{align}
so that the first constraint of the AdS$_2$ case in~\eqref{eq:Vconstr} is dropped. Recall however that we also require here that $\bar n_i=0$, so the AdS$_5$ solutions are actually much more strictly constrained than the AdS$_2$ solutions.  

Similar comments apply to the temperature dependence of this fixed point solution as the temperature dependence of the AdS$_2$ solutions above. The endpoints of the flowing vacuum solutions have $f(r)=1$. AdS$_5$ black hole solutions at generic $\rh$ in~\eqref{eq:fAdS5} are only linked to the flowing solutions in restricted temperature ranges: at high temperatures (for UV AdS$_5$ fixed points) or at low temperatures (for IR AdS$_5$ fixed points).

\begin{figure}[h!]
\center
\includegraphics[width=.45\textwidth]{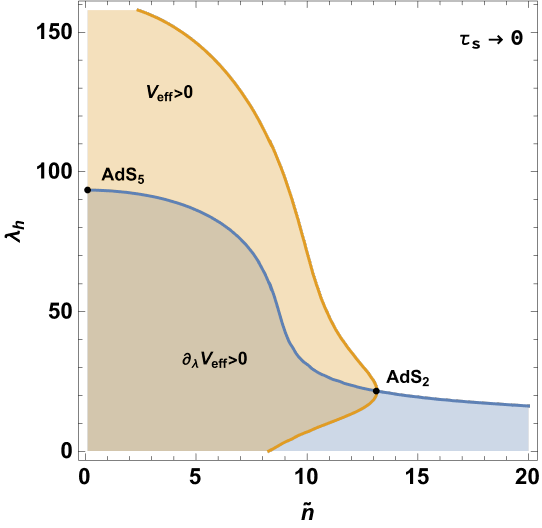}
\hspace{.05\textwidth}
\includegraphics[width=.45\textwidth]{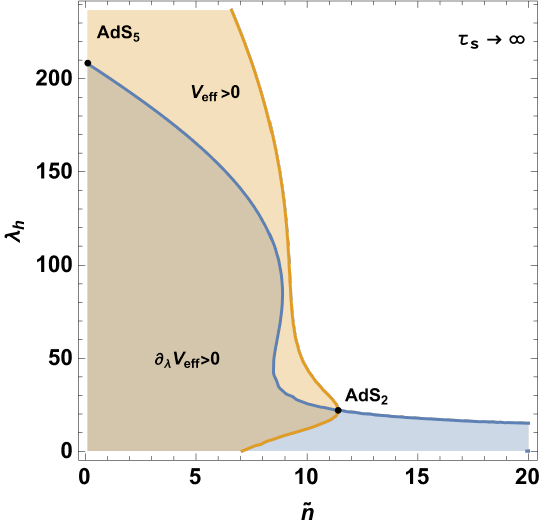}
\caption{The left and right plots on the top show the AdS$_2$ and AdS$_5$ fixed points for zero and infinite strange quark tachyon $\tau_s$ for the {``standard" potentials} ({\SP}) with the assumption that $\tilde{n}_{u,d} = \tilde{n}_s$.  The function form of tachyon dependence in $w(\lambda, \tau)$ is given in \eqref{eq:wtachyonform}. }\label{fig:fixedpoints}
\end{figure}

\begin{figure}[h!]
\center
\includegraphics[width=.45\textwidth]{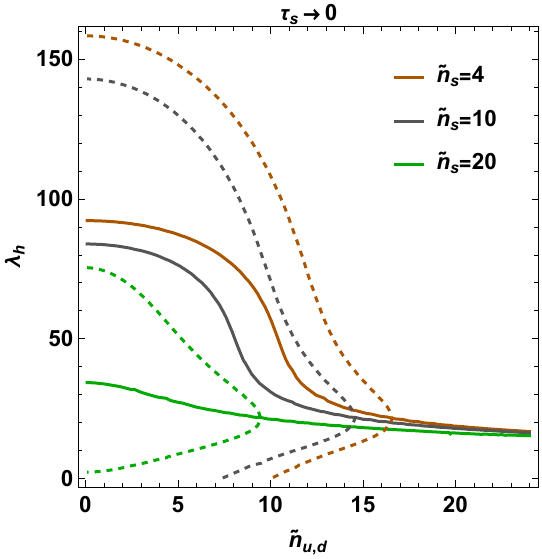}
\caption{The contour plot shows the AdS$_2$ and AdS$_5$ fixed points for different values of $\tilde{n}_s$ in the limit $\tau_s \rightarrow 0$. The solid curve shows the constraint $\partial_\lambda V_\mt{eff0} = 0$ and the dashed curve corresponds to $V_\mt{eff0} = 0$. 
}\label{fig:fixedpointsgen}
\end{figure}

\subsection{Fixed points realized in the V-QCD setup}

Above we identified possible AdS$_2$ and AdS$_5$ fixed points for the action we are using. However, which of these fixed points are realized in the model, and which of them appear in the physical phase diagram, depends on the choice of the various potentials in the action as well as the choice of quark masses. In the remaining of this article, we will focus on a $\Nf = 2+1$ setup (with two massless light quarks and one massive strange quark),\footnote{We often use a notation $i=u,d,s$ for the flavor indices, reflecting this setup, rather than $i=1,2,3$.} with potentials of the model adjusted to agree with lattice data. Then the following fixed points are relevant:
\begin{itemize}
    \item \textbf{The UV AdS$_5$ fixed point.} As usual in bottom-up holography, the trivial free theory fixed point of QCD at high energies is modeled by a zero coupling fixed point and an asymptotically AdS$_5$ geometry. This is achieved by the potential choices in V-QCD as the potentials have finite values, with power corrections in $\lambda$, in the weak coupling $\lambda \to 0$ limit~\cite{Gursoy:2007cb,Gursoy:2007er,Jarvinen:2011qe}. Apart from $\lambda_0=0$, also all tachyon components vanish, $\tau_{i0}=0$, at this fixed point. These values reflect the near boundary behavior for the fields for flowing solutions, see Appendix~\ref{app:asymptotics}. 
    Also all charges vanish, as required for AdS$_5$ fixed points.
    \item \textbf{The chirally symmetric IR AdS$_5$ fixed point.} The choices of potentials we use are such that the equation~\eqref{eq:laconstr} also has a solution at some nonzero $\lambda_0$ if all $\tau_{i0}$ vanish. The same fixed point also exists in the unflavored case~\cite{Jarvinen:2011qe,Alho:2012mh,Alho:2013hsa}. Because in this article we turn on a nonzero strange quark mass, it turns out that there are no flows which end on this fixed point.  However, there are flows which come close to this fixed point, so that it affects some of plots as we discuss below.  Note also that the Breitenlohner--Freedman (BF) bound is violated at the fixed point, see~\cite{Jarvinen:2011qe} and the analysis in Sec.~\ref{sec:ads5tau}, so flows ending on this point would be unstable towards condensation of some of the tachyon components. The value of the $\lambda_0$ for this fixed point solution is marked for a choice of potentials in the left panel of Fig.~\ref{fig:fixedpoints} where all quark densities are assumed to be equal.  
    \item \textbf{The chirally (partially) broken IR AdS$_5$ fixed point.} In this fixed point, the strange quark tachyon component diverges, $\tau_s \to \infty$, while the light quark components remain zero, $\tau_u=0=\tau_d$. We find that flows without spontaneous breaking of chiral symmetry can end in this point in the current model. However, these flows are unstable towards condensation of the light quark tachyons, indicating spontaneous chiral symmetry breaking in field theory, in the same way that the chirally symmetric AdS$_5$ point is unstable both here and in the unflavored case~\cite{Jarvinen:2011qe} due to the violation of the BF bound. It is marked in the right panel of Fig.~\ref{fig:fixedpoints}, again assuming equal densities for all quarks.
    \item \textbf{The chirally symmetric AdS$_2$ fixed point.} This is the generalization of the unflavored  chirally symmetric AdS$_2$ fixed point discussed in~\cite{Alho:2013hsa}. That is, all tachyon components vanish, $\tau_{i0}=0$, at the fixed point. It is marked in the left panel of Fig.~\ref{fig:fixedpoints} for the case where the densities of all quark flavors are equal, but it exists also for other ratios between the densities. As it turns out, this fixed point governs the zero temperature limit of the phase diagram. That is, at zero temperature the background flows are between the UV AdS$_5$ fixed point and this AdS$_2$ point. This happens even if we turn on a mass of the strange quark: tachyon perturbations around the fixed point are highly relevant, corresponding to a large dimension $\Delta_\mt{IR}$, so that $\tau_s$ rapidly vanishes as the fixed point is approached along the flow. We will discuss this behavior in more detail below. 
    \item \textbf{The chirally (partially) broken AdS$_2$ fixed point.} The fixed point equations also admit an AdS$_2$ solution where $\tau_s \to \infty$, while  $\tau_u=0=\tau_d$. This is marked in the right panel of Fig.~\ref{fig:fixedpoints}. {However, this solution does not seem to play any role in the final phase diagram of the V-QCD model. That is, it does not arise in any obvious limit from the numerical analysis of the solutions at nonzero temperature and density carried out in Sec.~\ref{sec:finitemu}.}
\end{itemize}

Finally, as we discussed in Sec.~\ref{sec:fixedpoints} the AdS$_2$ fixed points in the flavored case are more general than in the unflavored case, as the ratios of the quark densities may vary at the fixed point. We demonstrate this explicitly for the chirally symmetric fixed point in Fig.~\ref{fig:fixedpointsgen}, where we allow the (dimensionless) light quark densities $\tilde n_{u,d} = \tilde n_{u} = \tilde n_d$ to be different from the strange quark density $\tilde n_s$. Here the contours show where the conditions~\eqref{eq:Vconstr} and~\eqref{eq:laconstr} are satisfied so their intersection marks the AdS$_2$ fixed point. Interestingly, at large $\tilde n_s$ the AdS$_2$ fixed point disappears.

\subsection{Tachyon fluctuations around the AdS\texorpdfstring{$_5$}{5} point} \label{sec:ads5tau}

Let us then consider perturbing the fixed points. Fluctuations around the AdS$_5$~\cite{Jarvinen:2011qe} and AdS$_2$ points~\cite{Alho:2013hsa,CruzRojas:2024igr} in V-QCD have already been studied in detail in earlier literature. We only focus here on the results that are modified because of changes in the flavored setup with respect to the earlier unflavored studies. 

Tachyon fluctuations around the IR AdS$_5$ fixed points give an indicator of chiral symmetry breaking: if the BF bound is violated, the solution is unstable against tachyon condensation, which means spontaneous breaking of (some part of) chiral symmetry in the field theory. In V-QCD, the dimension related to the tachyon perturbation is given by~\cite{Jarvinen:2011qe}
\begin{align} \label{eq:DeltaIR}
    \Delta_\mt{IR}(4-\Delta_\mt{IR}) = -m_\mt{IR}^2 \ell^2_\mt{IR} \equiv \frac{24 a(\lambda_0)}{\kappa({\lambda_0}) V_{\mt{eff}0}}
\end{align}
where $m_\mt{IR}$ is the bulk mass of the tachyon at the fixed point and $\ell_\mt{IR}$ was evaluated by using Eq.~\eqref{eq:ellAdS5}.
The parameter $a(\lambda)$ appears from the flavor potential, $\Vf(\lambda,\tau_i) = V_{\mt{f}\lambda}(\lambda)\left[1-a(\lambda)\tau_i^2+\mathcal{O}\left(\tau_i^4\right)\right]$. Here we have fixed the normalization of the tachyon field such that $a(\lambda) = 1$. 

We compute here the IR mass of~\eqref{eq:DeltaIR} both for the chirally symmetric IR AdS$_5$ fixed point (where all $\tau_i \to 0$) and for the partially broken fixed point (where $\tau_{u,d} \to 0$ but $\tau_s \to \infty$) using the {``standard" potentials (\SP)} from Appendix~\ref{app:st_pots}. In the latter case, the mass refers to the fluctuation of the light tachyon components $\tau_{u,d}$.
The value of
$\lambda_0$ is determined by the constraints \eqref{eq:laconstr2}, setting 
$\hat{n}_i \rightarrow 0 $, 
\[
   \lambda_0 = \begin{cases}
       93.47, & \text{for } \tau_{i} \rightarrow 0 \\ 
       208.9, & \text{for } \tau_{u,d} \rightarrow 0, \  \tau_{s}  
       \rightarrow \infty 
        \end{cases}
  \]
which gives the IR AdS scale to be,

\[
   \ell^2_\mt{IR} = \begin{cases}
        0.2806, & \text{for } \tau_{i} \rightarrow 0 \\ 
       0.0714, & \text{for } \tau_{u,d} \rightarrow 0, \ \tau_{s} \rightarrow \infty
        \end{cases}
  \]
Finally, for the BF bound obtained from the relation~\eqref{eq:DeltaIR} we obtain 
\[
   \Delta_\mt{IR}(4-\Delta_\mt{IR}) = \begin{cases}
       7.332, & \text{for } \tau_{i} \rightarrow 0 \\ 
       4.700, & \text{for } \tau_{u,d} \rightarrow 0, \ \tau_{s} \rightarrow \infty 
        \end{cases}
  \]
That is, the BF bound is violated in both cases. This is expected, as the chiral symmetry is spontaneously broken in QCD.

\subsection{Tachyon fluctuations around the AdS$_2$ point} \label{sec:tauflucts}

As it turns out, the tachyon fluctuation around the AdS$_2$ fixed point changes drastically with respect to earlier analysis~\cite{Alho:2013hsa,CruzRojas:2024igr} due to the introduction of the tachyon dependence in $w$. The derivative of $w(\lambda, \tau_i)$ with respect to $\tau$ is numerically large at small $\tau_i$, modifying the tachyon equation drastically in the limit $\tau_i \rightarrow 0$.

For the (zero temperature) AdS$_2$ geometry, the blackening factor is given by
\begin{align}
    f(r) =  r^2 \frac{e^{2 A_0}}{6} \partial_{A_0} V_\mt{eff0} \ .
\end{align}
Inserting this in the tachyon equation of motion, we find that the tachyon perturbation satisfies, in the fully chirally symmetric case,
\begin{align}
    \ddot{\delta \tau_{i}} (r) + \frac{2}{r}\dot{\delta \tau_i} (r) + \frac{\Delta_i}{r^2}\delta \tau_i (r) = 0
\end{align}
where
\begin{align}
    \Delta_i = \frac{18}{\kappa(\lambda_0) \partial_{A_0} V_\mt{eff0}} \left( 1+ \frac{\tilde{n}_i^2}{V_{\mt{f}\lambda}(\lambda_0)^2 w(\lambda_0,0)^2} \right)^{-1} \left( 2+ \frac{\tilde{n}_i^2 ~ \left(\partial^2_{\tau_{i0}} w(\lambda_0,\tau_{i0})\right)\Big|_{\tau_{i0} = 0}}{V_{\mt{f}\lambda}(\lambda_0)^2 w(\lambda_0,0)^3}  \right)
\end{align}
where $V_{\mt{f}\lambda}(\lambda_0)=\Vf(\lambda_0,0)$.
The factor controlling the AdS$_2$ radius, $\partial_{A_0} V_\mt{eff0}$, 
reads
\begin{align}
    \partial_{A_0} V_\mt{eff0} = \frac{3}{\Nc} V_{\mt{f}\lambda}(\lambda_0) \sum_{j=1}^{3} \sqrt{1+ \frac{\tilde{n}_j^2}{V_{\mt{f}\lambda}(\lambda_0)^2 w(\lambda_0,0)^2}} \left(1 - \left(  1 + \frac{\tilde{n}_j^2}{V_{\mt{f}\lambda}(\lambda_0)^2 w(\lambda_0,0)^2} \right)^{-1} \right) 
\end{align}

The dimensionality $\Delta_{*i}$ reads as~\cite{Alho:2013hsa},
\begin{align} \label{eq:Ads2dimensionality}
    \Delta_{*i} = \frac{1}{2} + \frac{1}{2}\left(\sqrt{1-4 \Delta_i} \right)
\end{align}
which, for the standard potentials ({\SP}) in appendix~\ref{app:st_pots} and evaluated at the symmetric AdS$_2$ point, where $\tau_{i0} = 0$, gives $\Delta_{*i} = 5.566$ independently of the value of $i$. That is, the tachyon fluctuations at the fixed point are highly relevant, while in the absence of the tachyon dependence of $w(\lambda,\tau_i)$ the fluctuations were slightly irrelevant~\cite{CruzRojas:2024igr}, i.e., $\Delta_{*}$ was smaller than (but close to) one.

\section{Physics at small density} \label{sec:zeromu}

We then go on and discuss the background solutions, phase structure, and EOS of V-QCD at finite strange quark mass and small density. We will restrict to the case of zero light quark masses, i.e., $\Nf=2+1$ with two massless and one massive quark. For this analysis, we need to study backgrounds where the coupling and potentially some of the tachyons $\tau_i$ flow as a function of the holographic coordinate. We analyze the structure of such solutions and also comment on the relation to the fixed point solutions of Sec.~\ref{sec:fixedpoints}: They often cover the high and low energy limits of the flowing solutions. We will be mostly using the {\SP} given in Appendix~\ref{app:st_pots}, but also present some results for the alternative potentials (\AP) for comparison.

Most of the results are found by numerically constructing black hole solutions by shooting from the horizon. The method is a straightforward generalization of the method described in detail in~\cite{Alho:2012mh,Alho:2013hsa} to several independent tachyon fields. The values of $\Lambda$ and the quark masses are obtained by comparing the numerical solution to the near-boundary expansions given in Appendix~\ref{app:asymptotics}. The thermodynamic quantities are obtained by using the formulas given in Sec.~\ref{sec:thermo}.

\subsection{Background solutions at zero density} \label{sec:bgs_zero_density}

In order to draw the phase diagram, the first task is to identify all relevant classes of zero density background solutions. We discuss both thermodynamically stable and unstable phases. 

In the dual gravity, the relevant flowing solutions start from the AdS$_5$ UV fixed point at the boundary, and flow to different classes of geometries (including the fixed point solutions given in Sec.~\ref{sec:const_scalars}) or end in a horizon as one moves away from the boundary in the geometry, meaning lower energy in field theory. To be precise, the horizons and the fixed point solutions cover all possibilities for the end of the geometry except for one: a runaway geometry where the values of all scalars approach infinity. This gives rise to a geometry ending in a singularity, as we will discuss in more detail below.

As it turns out, the picture is analogous to that at zero strange quark mass~\cite{Alho:2012mh}, with minor modifications. That is, turning on the strange quark mass forces the condensation of the corresponding tachyon component $\tau_s$. This condensation suppresses the strange quark part of the flavor action at low energies. However, the light quarks behave as in the unflavored case discussed in~\cite{Alho:2012mh}: the tachyon fields $\tau_{u,d}$ may either condense, implying broken chiral symmetry for all quarks, or remain zero, implying that chiral symmetry remains intact in the light quark sector. Because of this, below we use terminology where chirally symmetric or broken solutions refer to the behavior in the light quark sector. The strange quark sector is always broken. As in the unflavored case~\cite{Jarvinen:2011qe}, the breaking of chiral symmetry is related to the violation of the BF bound for the light quark tachyon components at the AdS$_5$ IR fixed point (see Sec.~\ref{sec:ads5tau}).

We start by identifying the possible zero temperature solutions. These are found as solutions which flow from the AdS$_5$ UV geometry to special IR geometries as follows.
\begin{itemize}
    \item \textbf{Chirally broken vacuum.} This solution is a flow from the AdS$_5$ UV fixed point to a good IR singularity in the classification of Gubser~\cite{Gubser:2000nd}. In this case all tachyon components condense, approaching infinite values at the singularity, so the asymptotic singular geometry is exactly the same as in the unflavored case. Actually, since the whole flavor action is suppressed in the IR, it is also the same geometry as in the purely gluonic case, i.e., improved holographic QCD~\cite{Gursoy:2007cb,Gursoy:2007er}. However, the flavors affect the flow in the middle. This solution is the thermodynamically favored vacuum of the theory.
    \item \textbf{Chirally symmetric IR-conformal vacuum.} This is another zero temperature flow, running from the AdS$_5$ UV fixed point to the partially chirally broken AdS$_5$ IR fixed point. That is, the light quark tachyon components remain zero, while only $\tau_s$ condenses, so we call this solution chirally symmetric in our nomenclature. However, as it turns out, this solution is not thermodynamically preferred. Actually, it is perturbatively unstable: fluctuations of the light tachyon components violate the BF bound at the IR fixed point, see Sec.~\ref{sec:ads5tau}. The end point of this instability is the chirally broken vacuum. 
    \item \textbf{Higher order chirally broken ``Efimov'' vacua.} Apart from the ``standard'' chirally broken solution, there are also higher order chirally broken solutions~\cite{Jarvinen:2011qe}. The tachyon components for these solutions have nodes. All these solutions are highly unstable~\cite{Kutasov:2011fr,Arean:2013tja}. These vacua also appear in the unflavored case, and they are actually a rather generic consequence of the violation of the BF bound. However, in the flavored case the structure of the vacua is richer than in the unflavored case, because different components of the tachyon need to be treated independently.  We expect that the solutions for the three-flavor case with $m_s \ne 0$ may be labeled with $(n_d,n_u,n_s)$ where $n_i$ is the number of nodes\footnote{One could also include the uncondensed solutions in this classification following~\cite{Jarvinen:2024wsn}, but for simplicity we restrict here to vacua where all $\tau_i$ flow nontrivially.} for $\tau_i$. Due to symmetry, we may take $n_d \le n_u$. In this labeling $(0,0,0)$ is the ``standard'' chirally broken vacuum. 
\end{itemize}

We then go on and discuss the geometries at finite temperature. First there are the thermal gas solutions, which are trivially related to the zero temperature flows discussed above. That is, there are thermal gas versions of all chirally symmetric and broken vacua listed above, and the dependence of the background on the holographic coordinate is the same as for the zero temperature solutions. One can Wick rotate to Euclidean time and compactify the time coordinate, which leads to a regular finite temperature geometry. As in the case of the zero temperature solutions, only the standard chirally broken thermal gas solution is stable. The chirally broken thermal gas solution gives the description of the low temperature confined phase in the model, which has the chiral symmetry breaking pattern expected in QCD.

However, the most interesting finite temperature solutions for us are the black hole solutions, which describe the deconfined phases. Also for black holes, we find branches which are related to the zero temperature vacua. That is, the solutions are classified as follows:
\begin{itemize}
    \item \textbf{Chirally broken black hole branch.} There is a black hole branch (one-parameter family of solutions) for which all tachyon components are nonzero. In the limit of small black holes, the solution approaches the chirally broken zero temperature vacuum of the model. This behavior is required for the IR singularity of the vacuum to be of the good kind: the black hole horizon cloaks the IR singularity in the limit of small black holes~\cite{Gubser:2000nd}.
    \item \textbf{Chirally symmetric black hole branch.} There is also a branch of solutions for which the light quark tachyon components vanish, $\tau_u=0=\tau_d$. In the same way as for the chirally broken black holes, in the limit of small black holes this branch of solutions approaches the chirally symmetric vacuum (flowing to the AdS$_5$ IR fixed point). One can also analyze the limit of large black holes: Because of the near-boundary asymptotics of the tachyon field (see Appendix~\ref{app:asymptotics}) $\tau_s$ becomes suppressed for there black holes. Then the geometry approaches the fully uncondensed and chirally symmetric black hole. In the ultimate large black hole limit the geometry resembles the pure AdS$_5$ black hole, but the convergence is slow due to logarithmic corrections from the UV RG flow (these also appear in the expressions of Appendix~\ref{app:asymptotics}). The chirally symmetric black holes give the holographic model for the deconfined, chirally symmetric phase of QCD. 
    \item \textbf{Higher order black hole branches.} The violation of the BF bound at the AdS$_5$ IR fixed point also gives rise to higher order chirally broken black hole branches for which some of the tachyon components have nodes. However, solutions with several nodes only appear at exponentially small temperatures~\cite{Alho:2012mh,Alho:2013hsa}. All these black hole solutions are highly unstable. 
\end{itemize}

\begin{figure}[H]
\center
\includegraphics[width=.475\textwidth]{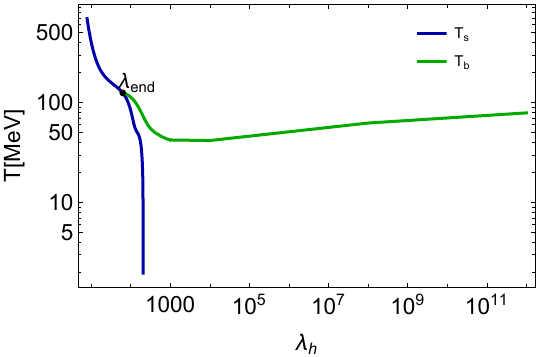}%
\hspace{.03\textwidth}%
\includegraphics[width=.495\textwidth]{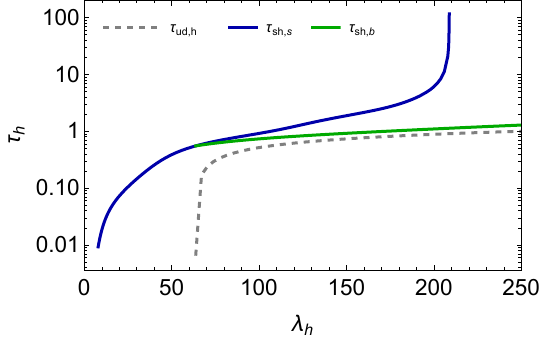}
\caption{Left: The temperature as a function of the horizon value $\lambda_\mt{h}$ of the broken and symmetric phase marked as $T_\mt{b}$ and $T_\mt{s}$. Right: the horizon values of tachyons for the black hole solutions as a function of $\lambda_\mt{h}$. The horizon value of the strange quark component is shown for both branches, while the light quark values are only shown in the broken branch, as they vanish in the symmetric branch.} \label{fig:3c}
\end{figure}

Let us then discuss the branch structure as a function of the black hole size and temperature. We show the temperature as a function of the horizon value of the dilaton, $\lah$, in Fig.~\ref{fig:3c}. We use here the standard potentials of Appendix~\ref{app:st_pots}, which were determined by fitting to lattice data as we will explain in Sec.~\ref{sec:lattice}. The value of $\lah$ encodes the size of the black hole such that larger $\lah$ means smaller black holes. This reflects the growth of the coupling in field theory as the energy decreases. 

We first focus on the chirally symmetric black holes, shown as the blue curve in the left panel of Fig.~\ref{fig:3c}. This branch shows the expected small black hole (low temperature) and large black hole (high temperature) limits: As $T \to 0$, $\lah$ approaches the AdS$_5$ fixed point value depicted in the right panel of Fig.~\ref{fig:fixedpoints}, and the geometry approaches the chirally symmetric vacuum geometry, flowing to the AdS$_5$ IR fixed point. As $T \to \infty$, $\lah$ approaches zero, and the geometry approaches the AdS$_5$ black hole geometry.

There is also a visible kink below $T=50$~MeV in the curve of the symmetric black holes in Fig.~\ref{fig:3c}. This kink arises from the interplay between different AdS$_5$ fixed point as follows. At high temperatures (small $\lah$), as the black holes are rather large, $\tau_s$ is a small perturbation all the way to the horizon. In particular, the horizon value of $\tau_s$, shown as the blue curve in the right panel of Fig.~\ref{fig:3c}, is suppressed in this region. Therefore the black hole geometry is well approximated by a fully symmetric solution, where $\tau_s=0$ also (so that $m_s=0$ as well). The black holes of this fully symmetric branch would approach at low temperatures the fully symmetric AdS$_5$ IR fixed point of the left panel in Fig.~\ref{fig:3c} instead of the partially broken fixed point of the right panel. However, when the temperature is lowered $\tau_s$ grows rapidly near the horizon (blue curve in the right panel in the figure), decoupling the strange quark sector.  This growth pushes the flow to the partially broken fixed point of the right panel, and the kink in the curve is a consequence of the slightly higher value of $\lambda$ at this IR fixed point.

Above we pointed out that the symmetric vacuum solution is unstable due to the violation of the BF bound for the light quark tachyon at the AdS$_5$ IR fixed point. This implies that the symmetric branch of black hole solutions is also unstable in the limit of small black holes. This can be seen concretely by considering a small homogeneous perturbation of the light quark tachyon components around the black hole background~\cite{Alho:2012mh}. In the zero temperature limit, the normalizable tachyon perturbation has an infinite number of nodes, signaling the instability and the violation of the BF bound at the fixed point. Conversely, in the high temperature limit the solution agrees with the power law solution given in the Appendix~\ref{app:asymptotics}, and has no nodes. Therefore as we move along the blue curve in Fig.~\ref{fig:3c} from low $T$ to high $T$, the nodes must disappear at some location. This happens at the point marked as $\lambda_\mt{end}$ in the plot, and it marks the location where the symmetric black hole solution becomes stable. 

Let us then study the chirally broken black hole branch, i.e., the green curve in Fig.~\ref{fig:3c}. In the limit of small black holes, which means $\lah \to \infty$ for this branch, the geometry approaches the chirally broken vacuum. Note that this holds even if the temperature of the black holes diverges logarithmically as $\lah \to \infty$, rather than approaching zero. As $\lah$ is lowered the black hole branch ends by merging with the symmetric branch at $\lah = \lambda_\mt{end}$, i.e., at the same point where the symmetric branch becomes stable. This is natural, as for $\lah>\lambda_\mt{end}$ the end-point of the instability of the symmetric branch is\footnote{At low temperatures, where the broken black hole branch does not exists, the end point is instead the broken thermal gas solution.} the broken branch. As the merging point is approached $\tau_{u/d}$ tend to zero. In particular, the horizon values of the light quark tachyon components approach zero as shown by the gray dashed curve in the right panel of Fig.~\ref{fig:3c}.  This will give rise to a second order chiral transition in the phase diagram~\cite{Alho:2012mh}. We also remark that at large $\lah$, where the black hole solution approaches the chirally broken vacuum, all tachyon components show similar behavior and the strange quark mass is merely a perturbation. This is demonstrated by the light quark (gray dashed curve) and strange quark tachyon horizon values showing the same large $\lah$ behavior in the right panel of Fig.~\ref{fig:3c}.
Moreover, note that only part of the chirally broken black hole branch is stable. At large values of $\lah$, the temperature increases with $\lah$, while the entropy decreases as the black hole becomes smaller. Therefore $ds/dT<0$, indicating a thermodynamic instability.

Finally, let us comment on the solutions at nonzero but small light quark mass. Turning on the mass is expected to remove the chiral transition arising from the branching point at $\lambda_\mt{h} = \lambda_\mt{end}$. As in the unflavored study~\cite{Alho:2012mh}, turning on a tiny quark mass, one finds a solution that follows the chirally symmetric zero mass solution for large black holes (blue curve at $\lah<\lambda_\mt{end}$ in the left panel of Fig.~\ref{fig:3c}) and the chirally broken solution for small black holes (the green curve blue curve at $\lah>\lambda_\mt{end}$ in the same panel), but is smooth near $\lah = \lambda_\mt{end}$.

\subsection{Thermodynamics at zero density} \label{sec:zeroTthermo}

Let us then discuss how the thermodynamics is determined for the black hole branches. The entropy and the temperature can be computed for each branch directly from the formulas~\eqref{eq:BHthermo}. However, the pressure or the free energy is more tricky to compute. {One can in principle compute the pressure by evaluating the on-shell action, after removing the UV-divergence via holographic renormalization \cite{Papadimitriou:2011qb}. However, implementing the holographic renormalization numerically in present case is technically challenging due to the presence of logarithmic terms in the near boundary expansion of the fields (see Appendix~\ref{app:asymptotics}) and the fact that the divergence of the action is severe.}\footnote{{It is inversely proportional to the fourth power of the UV cutoff if such a cutoff is introduced.}} The divergence disappears if one only considers pressure differences between phases, but even such differences are tricky to compute directly. Therefore, we use the fact that the pressure is known to be consistent with the thermodynamics of the black hole, and in particular satisfies the first law
\begin{equation} \label{eq:fstlaw}
    \dd P = s\, \dd T \ .
\end{equation}
Recall that all relevant solutions, including the broken and symmetric vacua, are connected through the black hole branches in Fig.~\ref{fig:3c}. Therefore, we can obtain the pressure differences between all solutions by integrating~\eqref{eq:fstlaw} along the branches shown in the plot. We will choose the constant of integration such that pressure vanishes for the chirally broken vacuum of the theory, which is obtained in the limit $\lah \to \infty$ along the broken branch (green curve) in Fig.~\ref{fig:3c}. 
For the confined thermal gas solutions all observables are temperature independent, and in particular the entropy vanishes, so that the thermodynamics is simple. That is, the only nontrivial number is the pressure, which equals the pressure of the corresponding vacuum solution.

\begin{figure}[H]
\center
\includegraphics[width=.6\textwidth]{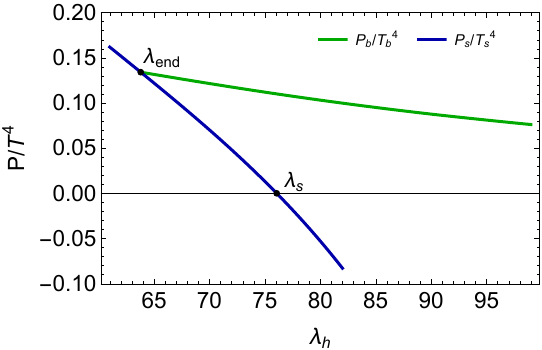}
\caption{The figure shows the dimensionless pressure as a function of the horizon value of $\lambda$. The $P_\mt{b}/T_\mt{b}^4$ and $P_\mt{s}/T_\mt{s}^4$ correspond to the pressure at the broken and unbroken phase. $
\lambda_\mt{end}$ marks the point where the pressure of both branches coincides, and $
\lambda_s$ marks the value of $\lambda $ near the horizon or the corresponding temperature where the pressure of the unbroken phase vanishes. }\label{fig:3d}
\end{figure}

\begin{figure}[H]
\center
\includegraphics[width=.6\textwidth]{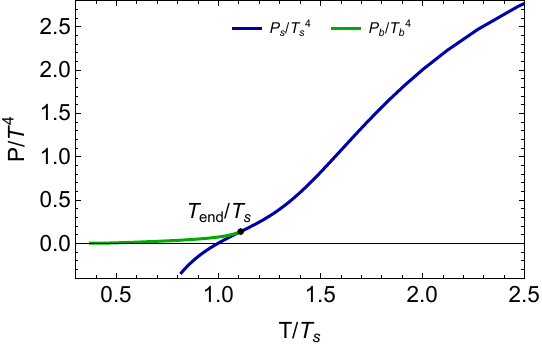}
\caption{The figure shows the dimensionless pressure as a function of $T/T_\mt{s}$. The $P_\mt{b}/T_\mt{b}^4$ and $P_\mt{s}/T_\mt{s}^4$ correspond to the pressure at the broken and unbroken phase. $
T_\mt{s}$ is the temperature at which the pressure of the symmetric branch vanishes. $T_\mt{end}$ is the temperature that marks the transition from the broken phase to the symmetric phase.} \label{fig:3e}
\end{figure}

We show the pressures obtained by integrating the first law~\eqref{eq:fstlaw} in Figs.~\ref{fig:3d} and~\ref{fig:3e}. Fig.~\ref{fig:3d} shows $P/T^4$ as a function of $\lah$ in the region where the two black hole branches meet. Note that as one starts to integrate~\eqref{eq:fstlaw} from $\lah = \infty$, where the pressure vanishes, the green curve first dips to negative values because $dT>0$ asymptotically along the curve as seen from Fig.~\ref{fig:3c}. Thereafter the sign changes, the pressure crosses zero, and is already positive when entering the range shown in Fig.~\ref{fig:3d}, so the phase dominates over the low temperature confined phase. We also mark the point $\lah=\lambda_s$ where the pressure of the symmetric black hole phase vanishes, i.e., it is equal to the pressure of the confined phase.

We show the normalized pressure $P/T^4$ as a function of the temperature in Fig.~\ref{fig:3e}. Here $T_\mt{s}$ is the temperature of the symmetric phase at $\lah=\lambda_s$ where the pressure vanishes. From here one can directly read off the zero-density phase diagram: As the temperature increases, there is a first-order deconfinement transition from the vacuum to the chirally broken black hole phase at $T \approx 42.1$~MeV, followed by a second-order chiral transition at $T\approx 124.4$~MeV. Note that the transition temperatures are low as compared to the crossover in QCD, as we shall see when comparing to data below. However, our interpretation is that the low temperature phase should be described as a hadron gas, which would arise from string loops on the gravity side~\cite{Alho:2015zua}, and therefore cannot be described by a classical geometry. Carrying out a stringy loop calculation is challenging, but one can replace the classical V-QCD results by a simple HRG model in the confined low temperature phase~\cite{Alho:2015zua,Demircik:2021zll}. We will leave an explicit construction of the complete model to future work, but compare our results to HRG models below in Sec.~\ref{sec:HRG}. Nevertheless, the details at low temperature in Fig.~\ref{fig:3e} will be insignificant in the final construction.

\subsection{Comparison to lattice data}\label{sec:lattice}

We then discuss how the potentials $\Vg$, $\Vf$, $w$, and $\kappa$ are pinned down by using lattice data. As explained in Sec.~\ref{sec:setup}, we use as the starting point the parametrizations and fits presented in~\cite{Jokela:2018ers}. Specifically, we use the potentials 7 and 9 given in this reference. Here potentials 7 has become a standard reference choice close to the center of the parameter space, used for example as an intermediate choice in~\cite{Ecker:2019xrw,Demircik:2021zll}. However, as it turns out, we obtain a phenomenologically more favorable EOS at high density by starting from potentials 9. We will coin the new fit based on potentials 9 as the ``standard'' potentials (\SP) and the fit based on the old potentials 7 as the ``alternative'' potentials (\AP). See Appendix~\ref{app:st_pots} for precise definitions.

In the unflavored fit of~\cite{Jokela:2018ers}, the fit could be carried out in stages. First, the only potential of the gravity sector, $\Vg$, was fitted to data for pure glue, i.e., the EOS of Yang-Mills theory in the deconfined phase. The comparison to full QCD data was done also in the deconfined phase, keeping quark masses at zero so that the model was chirally symmetric.
In the second step, $\Vf$ was fitted to match the EOS of full QCD (using lattice data with 2+1 flavors) at zero density, which only depends on $\Vg$ and $\Vf$ in the model. In the third step, $w$ was determined by fitting the data for baryon number susceptibility. Due to working in the chirally symmetric setup, the coupling function of the tachyon $\kappa$ did not affect the thermodynamics of the deconfined phase. However, it did affect the fit indirectly since the pressure difference between the confined, chirally broken phase and deconfined phase did depend on $\kappa$. However, due to the simplicity of this dependence it was easy to handle it separately. 

Similar strategy still works in the flavored case with nonzero strange quark mass. In particular, we choose not to change $\Vg$ at all and use directly the function given in this reference, and only fit the potentials in the DBI action. However there are small complications in this fit: First, due to turning on explicitly the strange quark mass, the results for the EOS and the susceptibilities now directly depend on the tachyon coupling $\kappa$. However, this dependence turns out to be so weak that the strategy of the unflavored fit still works, i.e., the dependence is negligible as compared to the uncertainties in the lattice data. Second, since the strange quark tachyon field is nonzero, we need to worry about the tachyon dependence in the potentials. This will be constrained among other things by the difference in the light quark and strange quark number susceptibilities, as we explain below.
In the fits to full QCD data, we only fit the results in the deconfined phase (chirally symmetric black holes) to the data in the deconfined region, i.e., $T \gtrsim 150$~MeV. As the idea will be to use a HRG model for $T \lesssim 150$~MeV in the final construction, it makes no sense to fit the model in this region. That is, we are following the approach suggested in~\cite{Jokela:2018ers}. Note that much of the literature which uses the Maxwell action for vectorial gauge fields, follows a different strategy, where QCD data is fitted at all temperatures~\cite{Gubser:2008ny,DeWolfe:2010he,Knaute:2017opk,Critelli:2017oub,Cai:2022omk,Fu:2020xnh,Jokela:2024xgz}. See also~\cite{Misra:2019thm,Kushwah:2024ngr} for progress using a top-down approach.

We proceed to determine the potentials by comparing the model to lattice data with $2+1$ flavors at $\mu=0$. We observe that a finite strange quark mass in the model requires dependence of the potential $w(\lambda, \tau_i)$ on the tachyon field, while $\kappa(\lambda)$ remains independent of tachyon as it only affects the location of the phase transition in V-QCD from confined to deconfined phase. The gluonic potential remains unchanged and retains the same form defined in the earlier literature with zero quark mass. The flavor potential $\Vf$ changes, demanding a slow decay of tachyon in $\Vf$ similar to the baryonic fit in \cite{Jarvinen:2022gcc}.

\begin{figure}[H]
\center
\includegraphics[width=.485\textwidth]{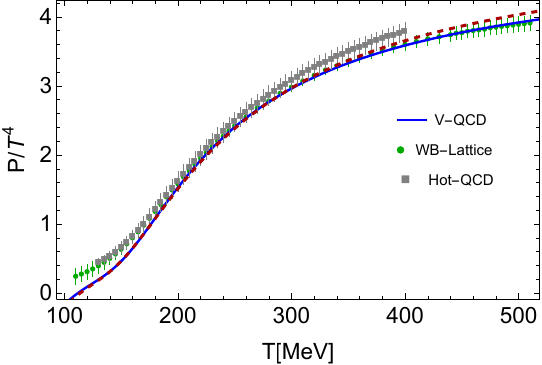}%
\hspace{.03\textwidth}%
\includegraphics[width=.485\textwidth]{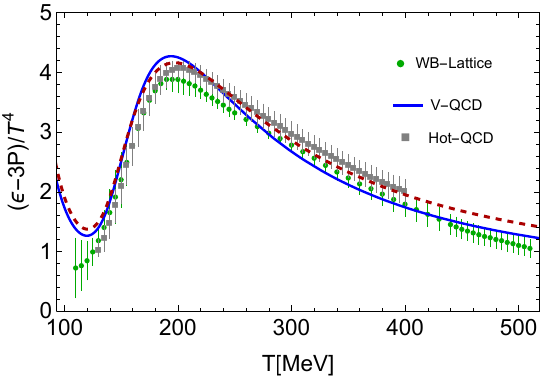}
\caption{The figure compares the pressure (left panel) and the interaction measure (right panel) computed in V-QCD model with the WB-lattice data~\cite{Borsanyi:2013bia} and the HotQCD data~\cite{HotQCD:2012fhj} of the EOS of $(2+1)$ flavor QCD. The blue V-QCD curve corresponds to the {\SP}, and the dashed red curve corresponds to the \AP.}\label{fig:3b}
\end{figure}

We start with the fit of pressure and interaction measure of QCD to lattice data (see Fig.~\ref{fig:3b}). Because the gauge fields vanish in the zero density background, this fit is independent of the shape and tachyon dependence of $w(\lambda, \tau_i)$, but depends on the strange quark mass. We observe that with increasing strange quark mass, it becomes difficult to fit the interaction measure and its integral $P/T^4$ simultaneously due using the Ansatz from~\cite{Jokela:2018ers} to a dip in the interaction measure in the region around or below $120$~MeV: The dip is due to the rapid growth of the strange quark tachyon component in the IR, so that the prediction of the V-QCD model lies significantly below the lattice data. Since we do not attempt a precise fit to lattice data in this region, this discrepancy is not an issue per se. However, $P/T^4$ is an integral of the interaction measure, and it turns out that agreement with lattice data for $P/T^4$ in the region $T \gtrsim 150$~MeV is only possible if the dip in the interaction measure is removed. Therefore, to ensure a dip-free fit, 
we make the following changes. Firstly, we set the strange quark mass to a lower value than in~\cite{CruzRojas:2024etx}, $m_s/\Lambda = 0.2839$, because a non-zero quark mass introduces non-linearities in the tachyon, which enhances with increasing mass, making precise fitting of the data impossible at low temperature. Secondly, the flavor potential $\Vf$ is modified as~\cite{Jarvinen:2022gcc}
\begin{align} \label{eq:Vfansatz}
    \Vf(\lambda, \tau_i) = \Vf{_\lambda}(\lambda) e^{-\tau_i^2}\left( 1+ \tau_i^4\right)^{\tau_p}
\end{align}
where choosing $\tau_p=0.6$ makes the potential gentler (than $\tau_p=0$) and reduces the growth of the tachyon in the IR, removing the dip from the interaction measure fit. Lastly, to ensure that the  temperature where the pressure of the symmetric branch vanishes in the holographic model,  $T_\mt{s}$, is a relatively high number, above $110$~MeV, we also fix the parameter $\bar{\kappa}_1$ of $\kappa(\lambda)$ (see Appendix~\ref{app:st_pots}) to a value higher than usually used in other potential classes of this model \cite{Jokela:2018ers}. 
As we discuss in Sec.~\ref{sec:finitemu}, this choice improves the consistency between the holographic quark matter EOS  and nuclear matter EOSs.

\begin{figure}[H]
\center
\includegraphics[width=.6\textwidth]{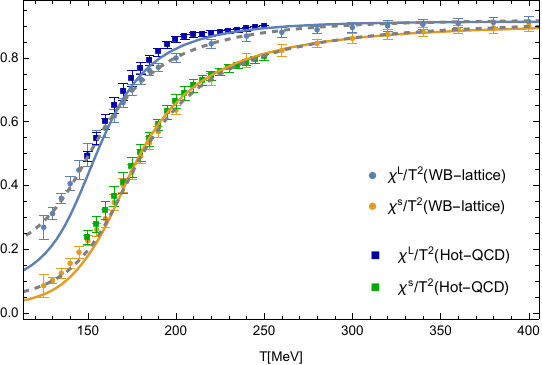}
\caption{The plot compares the light quark and strange quark susceptibility as a function of temperature with the WB-lattice data~\cite{Borsanyi:2011sw} and the HotQCD data~\cite{HotQCD:2012fhj} for {\SP} (solid curves). The dashed curve corresponds to the \AP.}\label{fig:3a}
\end{figure}

With the new set of parameters and tachyon dependence of $\Vf$, we proceed to fit $w(\lambda)$ to the light quark susceptibility, followed by the strange quark susceptibility by including tachyon field dependence in $w(\lambda, \tau_i)$ (see Fig.~\ref{fig:3a}).\footnote{We computed the shown HotQCD quark number susceptibilities and error estimates using the baryon number, charge, and strangeness susceptibilities reported in~\cite{HotQCD:2012fhj}.} The qualitative analysis of the lattice data suggests that the form of $w(\lambda, \tau_i)$ as a function of tachyon field
\begin{align}
\label{eq:wtachyonform}
    w(\lambda, \tau_i) =  \frac{w_\lambda(\lambda)}{1+ \beta_s \tanh({\gamma_s^2 \tau_i^2})}
\end{align}
where $\beta_s$ and $\gamma_s$ are arbitrary parameters set to $(\beta_s=0.65, \gamma_s=10 ~ \text{for standard potential})$ for an optimal fitting and $w_\lambda(\lambda)$ parameters are given in  Appendix~\ref{app:st_pots}. Note that when evaluated on the solution, $\Vf$ and $w$ depend on the tachyon only via the strange quark component $\tau_s$ as we only have non-zero strange quark mass. 

The result of the light quark susceptibility for the {\SP} in Fig.~\ref{fig:3a} undershoots the lattice data in the region of small temperatures, $T \lesssim 150$~MeV, where we do not fit the model to data. This is intentional: similarly as the choice of the function $\kappa(\lambda)$, choosing $W(\lambda)$ such that it produces a low susceptibility at low temperatures improves the consistency of the high-density EOS with nuclear matter EOSs. We discuss this more in Sec.~\ref{sec:finitemu}.

We remark that the fit to the strange quark susceptibility is not possible without a strong tachyon dependence, i.e., high value for the parameter $\gamma_s$, in the function $w(\lambda,\tau_i)$. In principle the tachyon dependence in $\Vf$ of~\eqref{eq:Vfansatz} already induces a difference between light quark and strange quark susceptibilities. However, the difference is smaller by orders of magnitude than the effect seen on the lattice. It could be enhanced by increasing the strange quark mass, but this would lead to the temperature dependence of the effect being incorrect, and also would make the fit to the pressure and to the interaction measure complicated. Therefore, adding a strong tachyon dependent correction such as that of~\ref{eq:wtachyonform} seems to be the only way to fit the lattice data.

Note that in addition to the light quark and strange quark susceptibilities lattice data~\cite{Borsanyi:2011sw,HotQCD:2012fhj} also shows nondiagonal, cross-flavor susceptibilities which are consistent with zero in the high-temperature deconfined phase but become nonzero (while still being suppressed with respect to the diagonal elements) near the crossover. However, in our holographic model, such nondiagonal susceptibilities vanish, unless one moves to nonzero density. Therefore it is not possible to precisely fit the full susceptibility data with our model. A possible interpretation is that this is not a shortcoming of the model, but reflects the fact that the holographic model only gives a valid description of deconfined quark matter, while the nondiagonal susceptibilities signal the transition to confined hadronic matter. 

Importantly, our choice is different from the choice done in all earlier works where the holographic model was fitted to lattice data for susceptibilities (see, e.g.~\cite{DeWolfe:2010he,Knaute:2017opk,Critelli:2017oub,Cai:2022omk,Fu:2020xnh,Jokela:2024xgz}).\footnote{Note that apart from using the baryon number susceptibility, many of these fits have the weakness that the Ansatz for the function corresponding to our $w(\lambda)$ produced an unphysical spike at small $\lambda$. See~\cite{Jokela:2024xgz,Lilani:2025wnd} for discussion of this issue.} Namely, the baryon number susceptibilities were used, which actually depend on both on the diagonal and nondiagonal quark number susceptibilities. Unlike these earlier fits, the fit carried out in this article does not give a precise description of the baryon number susceptibilities near the crossover. However, it seems to us that our choice here is the most natural way to compare to the full flavor dependence in the data.

\subsection{Comparison to HRG} \label{sec:HRG}

As we mentioned above, our aim is to use the fit established in this article to construct a hybrid model where, in the region of low baryon number density, holography is used at temperatures above the crossover and HRG below it~\cite{Alho:2015zua,Demircik:2021zll}. We do not attempt to construct such a hybrid explicitly here, but 
we compare pressure over temperature 
from V-QCD with those of HRG models at zero strangeness and chemical potential. 

\begin{figure}[H]
\center
\includegraphics[width=.6\textwidth]{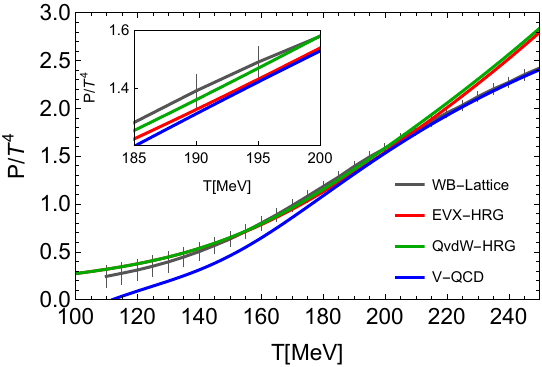}
\caption{The figure compares the quantity $P/T^4$ as a function of $T$ for different HRG models with the V-QCD model in blue.}\label{fig:6}
\end{figure}

We compare the V-QCD result with those from the excluded HRG model \cite{Rischke:1991ke,Vovchenko:2020lju}, and the quantum van der Waals HRG \cite{Vovchenko:2015vxa, Vovchenko:2016rkn} using the THERMAL-FIST package \cite{Vovchenko:2019pjl}. With this package, we generate 
the thermodynamic quantities for different HRG models based on the interaction parameters given in the literature \cite{Vovchenko:2016rkn} for zero chemical potential and zero strangeness. Figure \ref{fig:6} compares $P/T^4$ computed from the V-QCD model with the non-diagonal HRG model (EVX) and the quantum van der Waals (QvdW) HRG model. The inset plot shows that the EVX-HRG model to a good precision overlaps with the V-QCD model around $T \sim 180-220$ MeV.

\section{Nonzero density and small temperature}\label{sec:finitemu}

We then discuss turning on a nonzero chemical potentials for the quarks, focusing on the physics at small temperature. We keep again nonzero the strange quark mass while setting the light quark masses to zero.

\subsection{Background solutions at finite density}

First we review the classes of geometries at nonzero chemical potential, in analogy to the discussion of Sec.~\ref{sec:bgs_zero_density}. 

To start with, we note that all vacua and thermal gas solutions listed in Sec.~\ref{sec:bgs_zero_density} admit a trivial generalization to finite chemical potential but zero  number density. These are obtained by turning on a constant temporal components for the gauge fields, $\Phi_i$ in Sec.~\ref{sec:thermo}, which are therefore equal to the quark chemical potentials for various flavors. Nonzero charges $n_i$ are disallowed: thermal gas solutions with $f(r)=1$ are only possible at zero charge, see the equations of motion in Appendix~\ref{app:EOM}.

However, at nonzero density, there is a class of solutions that is new and distinct from the zero density solutions, i.e., a new geometry at zero temperature:
\begin{itemize}
    \item \textbf{Chirally symmetric nonzero density vacuum.} This solution flows from the AdS$_5$ UV fixed point to the chirally symmetric AdS$_2$ fixed point listed in Sec.~\ref{sec:fixedpoints}. As above, ``chirally symmetric'' only refers to the behavior in the light quark sector, even if the fixed point is fully chirally symmetric. The strange quark component $\tau_s$ is condensed due to the nonzero strange quark mass. However, as we discussed in Sec.~\ref{sec:tauflucts}, $\tau_s$ is suppressed at the AdS$_2$ fixed point because tachyon fluctuations are this fixed point are highly relevant. We do not study these flows in detail in this article, but they appear to be simple generalizations of the unflavored flows studied in~\cite{Alho:2013hsa,Hoyos:2021njg}. That is, the total charge is fixed in units of entropy as in the case of the unflavored flows, but flavor dependence allows us to adjust the ratios of the densities of different quark flavors. 
\end{itemize}
Moreover, the chirally symmetric IR-conformal vacuum (see Sec.~\ref{sec:bgs_zero_density}) is not present at nonzero density as the AdS$_5$ IR fixed point only exists at zero density. 

As for the black hole solutions, all classes (including chirally symmetric and broken branches) admit generalizations to nonzero density. While we do not attempt to analyze the limit of small black holes in detail here, we expect that the symmetric and broken black hole solutions approach the symmetric and broken vacua at nonzero chemical potential, respectively, as is the case for the unflavored solutions~\cite{Alho:2013hsa}. In other words, the vacua satisfy the consistency conditions~\cite{Gubser:2000nd} so that in particular the broken IR singularity remains to be of the good kind also at nonzero density.

\subsection{Thermodynamics at finite density (and small temperature)}

Let us then discuss thermodynamics of the various phases at nonzero density. The analysis is a generalization of the zero density analysis of Sec.~\ref{sec:zeroTthermo}. 

The relevant geometries to study are the chirally broken thermal gas solution, which models the confined phase of QCD, and the chirally symmetric black hole solution, which models the deconfined phase. As we saw in Sec.~\ref{sec:zeroTthermo}, also the chirally broken black hole phase appears in the nonzero density phase diagram of the model, so it will also be present among the nonzero density phases at least at small values of the densities. However we do not attempt to analyze this phase at nonzero density in this article. As we saw in Sec .~\ref {sec:HRG}, at zero density the pressure of the broken deconfined phase was far from giving a reasonable model for the pressure of real QCD, and we expect the same to hold at nonzero density.

Since the gauge fields of the thermal gas solutions are constants, they do not contribute to the on-shell action. Consequently, the pressure of these solutions is independent of the chemical potentials and equal the pressure of the zero density thermal gas vacuum. However, the black hole solutions have highly nontrivial thermodynamics. As above, we can compute the entropy, temperature, densities and chemical potentials directly from the numerically constructed black hole solution by using the formulas from Sec.~\ref{sec:thermo}. The pressure is then obtained by integrating the first law,
\begin{equation} \label{eq:firstlawfull}
   \dd P = s\, \dd T + \sum_{i=1}^{\Nf} n_i \dd \mu_i \ .
\end{equation}

We analyze here chirally symmetric black holes at high density and low temperature. We carry out the analysis at $T=5$~MeV: this gives a good approximation to thermodynamics at zero temperature, and is easier to do than the zero temperature analysis (which would require studying numerically the flows from the UV AdS$_5$ fixed point to the IR chirally symmetric AdS$_2$ fixed point). Rather than exploring the full parameter space, we wish to compute the thermodynamics as a function of a single chemical potential (the baryon number chemical potential), but since there are several quark chemical potentials, it is not immediately obvious what is the best choice for fixing them. We choose here to impose the following conditions: We require the plasma to be charge neutral, $2n_u-n_d-n_s=0$, and require chemical equilibrium for the down type quarks, $\mu_d = \mu_s$. These conditions correspond to $\beta$-equilibrated matter in the absence of leptons. These requirements can be met by writing $n_d=n_u+\delta n$, $n_s=n_u-\delta n$ and adjusting $\delta n$ until  $\mu_d = \mu_s$ is obtained. Typically $\delta n \ll n_u$ so that the quark densities are almost equal. We show our results in terms of the quark chemical potential $\mu=(\mu_u+\mu_d+\mu_s)/3$.

However there is a complication: the family of chirally symmetric black holes at fixed $T=5$~MeV is not connected to the zero density black hole solutions we considered in Sec.~\ref{sec:zeroTthermo}. This means that the integration constant remains unfixed as the pressure is solved from~\eqref{eq:firstlawfull}. In order to determine the integration constant, we need to connect the high-density black hole solutions to the zero density black hole solutions. We do this by constructing numerically another branch of black holes which lie on an arc on the $(\mu,T)$ plane. Specifically, this arc is part of an ellipse (centered at $\mu=0$ and $T=5$~MeV) which starts from $\mu=0$, $T=155$~MeV and ends at $\mu=500$~MeV, $T=5$~MeV. The integration constant can then be determined by integrating the pressure along this branch, and requiring that the pressure obtained in the limit of zero $\mu$ agrees with the pressures computed in Sec.~\ref{sec:zeroTthermo}.

\begin{figure}[H]
\center
\includegraphics[width=.8\textwidth]{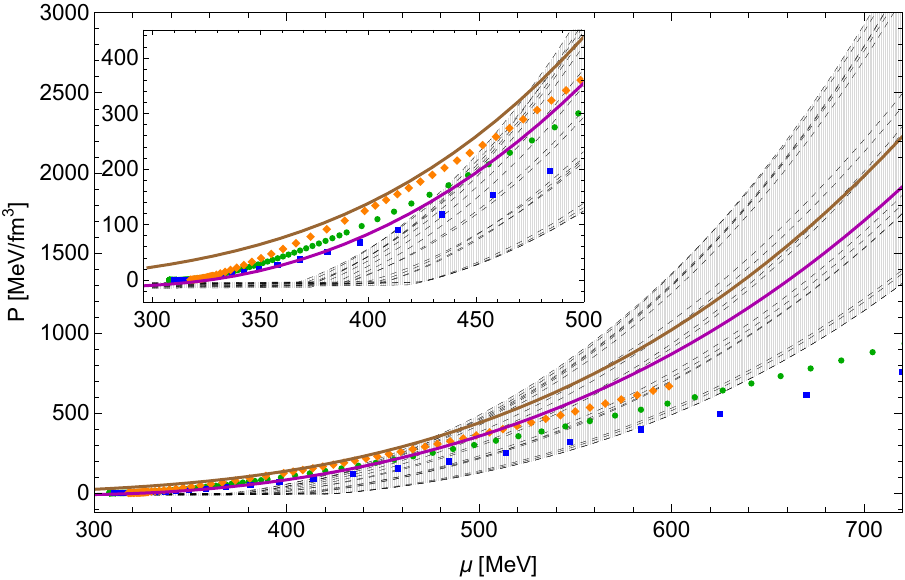}
\caption{The figure compares the pressure at zero temperature of the chirally symmetric black holes as a function of quark chemical potential for different flavor potentials and nuclear theory models. The solid magenta (solid brown) curve shows the pressure for {\SP} (\AP) at non-zero strange quark mass. The black dashed curves and the gray band show the pressure for different potentials with zero strange quark mass discussed in \cite{Jokela:2018ers}. The dots show results from nuclear theory models: The green dots are the APR data, the blue are DD2, and the orange are the soft HLPS variant.}\label{fig:5}
\end{figure}

We plot the pressure at zero temperature\footnote{To be precise, we plot $P-sT$ with all quantities evaluated at $T=5$~MeV, which is the linearized approximation for the pressure at zero temperature. The difference between the pressure at $T=5$~MeV and this zero temperature approximation is so small that would be barely visible in the plots.} of the chirally symmetric black holes, i.e., the deconfined quark matter phase, as a function of the quark chemical potential for the potentials of Appendix~\ref{app:st_pots} in Fig.~\ref{fig:5}. The result for the {\SP} (solid magenta curve) and {\AP} (solid brown curve) are compared to the families of equations of state constructed using the unflavored model in~\cite{Jokela:2018ers} (black dashed curves and the gray band) and a couple of nuclear theory models (colored dots). The green dots, blue boxes, and orange diamonds are given by the Akmal-Pandharipande-Ravenhall (APR) model~\cite{Akmal:1998cf}, the DD2 variant of the Hempel-Schaffner-Bielich model~\cite{Hempel:2009mc,Typel:2009sy}, and the soft variant of the Hebeler-Lattimer-Pethick-Schwenk (HLPS) model~\cite{Hebeler:2013nza}, respectively. Remarkably, new results match with the nuclear theory models much more smoothly than any of the old unflavored models  show as black dashed curves. We stress that this happens even when none of the parameters of the potential are adjusted to any observables at high density. The difference with respect to the earlier fits is mostly due to the increase of $\tau_s$ deactivating the strange quark sector, which tends to reduce the values of the chemical potential.

However we note that the result for the {\AP} does not intersect the data for any of the nuclear theory models (even though it comes extremely close), but the V-QCD pressure is above the nuclear matter curves for all shown chemical potentials. This is not a phenomenologically desired behavior: in the end we want to construct a model which also includes nuclear matter, modeled either by using holography or by some other framework, and the intersection between the nuclear matter and V-QCD pressure would mark the phase transition from nuclear to quark matter. That is, using the {\AP} do not seem to give rise to a reasonable phase diagram in this region. This was actually our motivation for introducing the other set of potentials, i.e., the \SP. Importantly, these potentials have a higher value of the parameter $W_0$ which controls the normalization of the DBI action at small $\lambda$, and is not determined by the comparison to lattice data for QCD thermodynamics~\cite{Jokela:2018ers}. Basically, the fit has a flat direction, which is parametrized using this parameter. As it turns out, increasing $W_0$ slightly lower the result for the pressure at high densities, helping us to obtain a phenomenologically favored behavior. Apart from adjusting the value of $W_0$, we also tuned other parameters that are not directly controlled by the lattice fit to improve the behavior of the pressure in the region plotted in Fig.~\ref{fig:5}: We chose the function $\kappa(\lambda)$ such that the pressure difference between the confined and deconfined phases was relatively high, and chose the function $w(\lambda)$ at large values of $\lambda$, where it no longer affects the quark number susceptibilities, to be as small as possible. Both these have the effect of slightly lowering the V-QCD pressure with respect to the nuclear matter models. The need of such adjustments in order to obtain a phenomenologically viable EOS means that (unlike in the unflavored case) there are strong additional constraints to the parameter space of the model from the high-density behavior. After taking into account these constraint, the final EOS from the model is determined significantly more precisely than in the earlier unflavored studies.

\begin{figure}[H]
\center
\includegraphics[width=.6\textwidth]{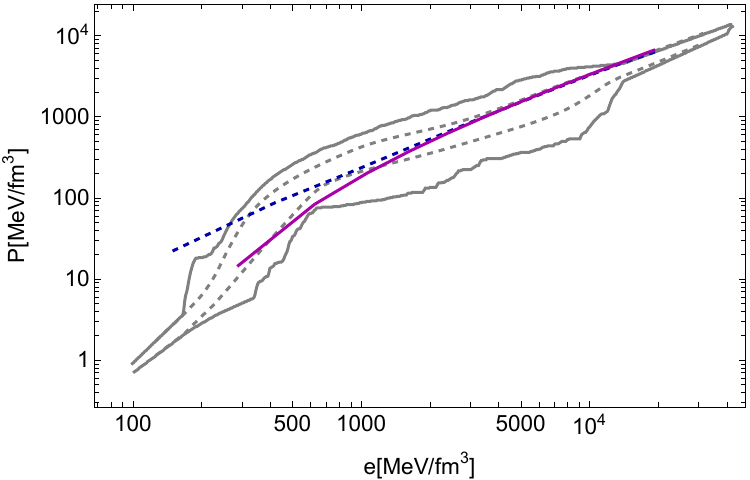}
\caption{The figure shows the logarithm of the pressure as a function of the logarithm of the energy density for {\SP} (solid magenta curve) and {\AP} (dashed blue curve) compared to bounds from the model-independent approach of~\cite{Ecker:2022dlg} (gray dashed and solid curves). 
}\label{fig:sausage}
\end{figure}

We compare our EOS to general model-independent constraints, coming from interpolations between nuclear theory and perturbative QCD predictions in Fig.~\ref{fig:sausage}~\cite{Ecker:2022dlg}. The result for the {\SP} (solid magenta curve) stays within the 100\% band (gray solid curves) even down to the low density nuclear matter region, which reflects the smooth matching of the nuclear matter models. However, the result for the {\AP} exits the band from the top at very low densities, which indicates a problem with matching, in agreement with the comparison to nuclear matter models in Fig.~\ref{fig:5}.

\begin{figure}[H]
\center
\includegraphics[width=.5\textwidth]{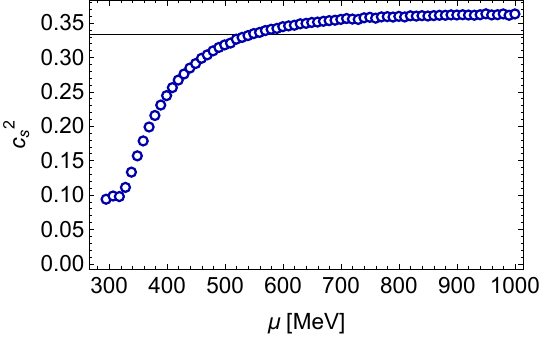}
\caption{The figure shows the speed of sound as a function of quark chemical potential at low temperature 5 MeV. The black horizontal line shows the conformal limit of the speed of sound. }\label{fig:4}
\end{figure}

We also wish to study the adiabatic speed of sound 
\begin{equation}
    c_s^2 = \frac{\partial P}{\partial \epsilon}\Big|_{\tilde n_i\ \textrm{fixed}} \ ,
\end{equation}
where $\tilde n_i = 4\pi n_i/s$.
However, rather than computing the precise physical speed of sound, we compute the quantity corresponding to the EOS at fixed $T=5$~MeV, with charge neutrality and $\mu_d=\mu_s$, given by
\begin{equation}
     c_s^2 \approx \frac{d \log \mu}{d \log n} 
\end{equation}
with the differential taken along the EOS curve, i.e., at constant temperature and at $\beta$-equilibrium, which is simpler to compute from our data.
We show the result as a function of the chemical potential in Fig.~\ref{fig:4}. This quantity is actually expected to be close to the physical speed of sound: the ratios of the densities $n_i$ between quark flavors are fixed equal to one up to corrections $\sim 1$~\%, and taking the temperature to zero fixes the overall $n/s$ also via the AdS$_2$ conditions~\eqref{eq:Vconstr} and~\eqref{eq:laconstr}. Similarly as in the unflavored case, the speed of sound exceeds the conformal value (horizontal black line in Fig.~\ref{fig:4}) at high densities.

\section{Conclusion} \label{sec:conclusion}

In this article, we constructed explicitly a flavor-dependent model for the thermodynamics of QCD at nonzero temperature and/or baryon number density. Specifically, we studied a version of the V-QCD model which significantly generalizes earlier variants given in the literature. We started by discussing a setup with general flavor dependence, i.e., a model with arbitrary number of flavors having different masses and chemical potentials. We identified the possible general constant scalar fixed point solutions of such a general setup, which may appear as end points of RG flows in the backgrounds of the theory at zero and nonzero temperature. 

We then focused on a construction with effectively 2+1 flavors, i.e., two massless light quarks and a massive (strange) quark. We draw the phase diagram of the setup, identified the relevant fixed point solutions, and finally adjusted the model action to agree with lattice data for thermodynamics in QCD in the deconfined quark-gluon plasma phase at low baryon number density. Moreover, we compared the model predictions at lower temperature to various hadron resonance gas models. Finally, we computed the EOS at high density and low temperatures, and compared to nuclear theory EOSs as well as constraints from model independent approaches, also taking into account neutron star and neutron star merger observations.

We remark that our fit to the lattice data for quark number susceptibilities was done in a slightly different way as earlier (unflavored) fits in the literature. Namely we fitted only the diagonal terms in the susceptibility matrix $\chi_{ij} = \frac{\partial^2 p}{\partial\mu_i\partial\mu_j}$ in Sec.~\ref{sec:lattice}, where $\mu_i$ are the chemical potentials for different flavors, while the unflavored fits used the baryon number susceptibility, which also depends on the nondiagonal terms. Therefore, the additional of the strange quark mass is not the only difference in the fine details of the EOS in our model with respect to the earlier models.

We found that the tachyon dependence of the unflavored model had to be modified in a rather specific way in order to obtain a good agreement with lattice data. Specifically, the slope of the tachyon in the DBI tachyon potential $\Vf(\lambda,\tau)$ had to be adjusted to be gentler than the simple exponential $\Vf \sim e^{-\tau^2}$ suggested by the rolling tachyon picture~\cite{Sen:2002in,Sen:2004nf} and string field theory analysis~\cite{Kraus:2000nj,Takayanagi:2000rz}. Interestingly, this is in agreement with the results of~\cite{Jarvinen:2022gcc}, where the unflavored model was fitted simultaneously to thermodynamics and meson spectra, and a good fit was also found to require a gentle tachyon potential. Moreover, we found that a rather strong tachyon dependence in the coupling function of the gauge field $w(\lambda,\tau)$ was required. 

Since turning on the strange quark tachyon forces as to consider potential functions depending on two variables $\lambda$ and $\tau$ when comparing to data, one might think that the parameter space is drastically increased, and that the fit presented here is only one out of many possibilities. However, it turns out that the effect of the strange quark mass in the deconfined phase is captured by rather small values of the tachyon field, so that our results are mostly only sensitive to the leading $\mathcal{O}(\tau^2)$ correction to the potentials at zero tachyon. This correction is strongly constrained by the comparison to lattice data. That is, all predictions in the deconfined phase are tightly constrained by the comparison to lattice data. 

One of our key results was the EOS at high density and low temperature. We found that this EOS can be matched nicely with nuclear matter EOSs at lower density only for a subset of potentials. For example, pressure for the alternative potentials was higher than nuclear matter models at around $\mu=350$~MeV and below, so that combining the pressures would lead to a peculiar phase diagram. Requiring that such behavior is absent, and the pressure stays below the nuclear matter results at low densities, therefore sets a constraint to our models. While we did not carry out a detailed scan over the parameter space in this article, it appears that at least a large value of the parameter $W_0$, which controls the normalization of the DBI action, is required. This is in contrast with the unflavored results, where comparison with nuclear matter EOSs did not lead to any constraints. Consequently, after taking into account the flavor dependence, the predictions from the model are much more tightly constrained than the earlier unflavored computations were suggesting.
Moreover, for those potentials for which matching with nuclear matter is possible (such as the standard potentials considered in this article) the matching is much more smooth than in the earlier unflavored fits. That is, the latent heat of the nuclear to quark matter transition is significantly lower than earlier.

Our work opens up a high number of various future directions to explore. First, the exploration of the phase diagram of the flavored V-QCD model as defined in Sec.~\ref{sec:setup} is not yet complete. We observed that there was a hairy, chirally broken black hole phase at low temperatures and zero chemical potentials. Even if we argued that this phase will be subdominant to meson gas, which would be described by loop corrections on the gravity side and therefore cannot be captured in the simple classical treatment of the current article, it would be important to map the boundaries of this phase at nonzero chemical potential. We do not expect that this chirally broken phase plays a role in the model at high density and low temperatures, but this should be checked. Another task would be simply carry out a more detailed scan of the chirally symmetric black hole solutions, starting from the $\beta$-equilibrium setup when both $T$ and $\mu$ are nonzero. One should compute the EOS  and check if the phase diagram is simple or if one obtains phase transitions between chirally symmetric black holes. When comparing the result with the hadron resonance gas, one should check if the picture for the critical end-point of the confinement-deconfinement transition suggested in~\cite{Demircik:2021zll,Ecker:2025vnb} is modified due to the inclusion of the strange quark mass.

Apart from carrying out a more detailed analysis of the phase diagram of the model fitted to QCD data here, one should also check if the data fit can be improved. 
Note that the comparison of QCD data here was restricted to lattice data for the EOS. At least in principle, it should be possible to extend this fit by also comparing the model (with the same potentials) to the data for two-flavor QCD~\cite{Fang:2015ytf,Zhao:2023gur}. Indeed, such data is available, albeit at high quark masses~\cite{Burger:2014xga}. This comparison can be done by taking $m_s \to \infty$, effectively removing the strange quark sector from the model of this article, and turning on a sizeable mass for the light quarks. But one should also compare the flavored model to other data, in particular properties of the confined phase vacuum such as meson spectra and decay constants. One could also study more systematically how the results depend on the choices of potentials in regions that are not directly constrained by lattice data, such as the coupling function $w(\lambda,\tau_i)$ of the gauge fields at large values of $\lambda$ or $\tau_i$. 

In addition to fine-tuning the model as defined in this article, one should also study the effect of additional terms of sectors in the action. An example is the inclusion of multi-trace terms in the flavor action~\eqref{eq:Sfgen}, i.e., terms involving more than a single trace over flavor indices. Such terms do not appear in the DBI actions obtained from string theory (see, e.g.,~\cite{Casero:2007ae}) as they are suppressed by factors of $1/N_c$ with respect to the single-trace terms and would correspond to loop corrections. This can be seen by comparing to the field theory diagrams in the 't Hooft limit in analogue to the treatment in chiral effective theory~\cite{Gasser:1984gg,Kaiser:2000gs}. However, in the Veneziano limit, which is the basis of the V-QCD model, these terms are not suppressed because the additional traces enhance the terms by factors of $N_f$. Such terms have not been considered in the V-QCD literature so far even in the unflavored case in part due to the added complexity. However, they could be a particularly interesting addition to the flavored setup, as they could add correlations between different flavors, potentially mending the difficulty in fitting the nondiagonal elements of the flavored susceptibility matrix discussed above.

Turning on the strange quark mass will also affect some of the recent studies of different phases than the unpaired quark matter phase in V-QCD. An example is the simple model for color superconducting paired phase constructed in~\cite{CruzRojas:2025fzs}.
Similarly, strange quark mass will modify the chirally broken vacuum which we did not directly construct numerically here. (That is, our numerical analysis covered chirally broken black holes that approach the vacuum solution in the limit of small black holes.)
As a first step, one could check how the change in the vacuum affects recently studied solutions for nucleons, constructed as solitons of the gauge fields in gravity~\cite{Jarvinen:2022gcc,Jarvinen:2022mys}, and the simple model for nuclear matter considered in~\cite{Ishii:2019gta}.

Another direction is to study more closely the model away from $\beta$-equilibrium (as implemented here) and isospin asymmetry, which has been studied intensively in the nuclear matter phase using the holographic approach recently~\cite{Kovensky:2021ddl,Kovensky:2021kzl,Bartolini:2022gdf,Bartolini:2025sag}. The most direct analysis is to study the symmetry energy of quark matter using the model, effectively extending the EOS to nonzero charge fraction, which has so far been done only though approximations in V-QCD~\cite{Chesler:2019osn,Demircik:2021zll}. Apart from isospin asymmetry, one should also check what happens when the density of strange quarks is much larger than the density of light quarks. We observed in Sec.~\ref{sec:const_scalars} that the IR AdS$_2$ fixed point disappears in this regime for standard potentials. Consequently, the geometry must change drastically at low temperatures and high densities. 
However, one can also study isospin asymmetry in the chirally broken vacuum background. This vacuum solution could act as an improved background for studying, e.g., isospin breaking effects in nuclear matter. 
A specific question to address in near future is the fate of the Nakamura--Ooguri--Park instability~\cite{Domokos:2007kt,Nakamura:2009tf,Ooguri:2010kt} in the flavored model. It was demonstrated in~\cite{CruzRojas:2024igr,Demircik:2024aig}, using the V-QCD model with zero quark masses, that this instability appears at surprisingly low values of $\mu/T$, in the region of the phase diagram potentially reachable by lattice analysis and heavy-ion collision experiments. However, it was also pointed out that the dependence on quark masses, and in particular the strange quark mass, is likely to be significant. Therefore, it is essential to repeat the instability analysis with the flavored model established in this article.

Yet another direct application of the model would be to update the predictions for the bulk viscosity from weak interactions coupled to QCD matter, which were originally computed by a simplified V-QCD setup in~\cite{CruzRojas:2024etx}, possibly taking into account refined formulation from~\cite{Hoyos:2024pkl,Hernandez:2025zxw}. That is, we did turn on a nonzero strange quark mass as required to obtain a meaningful, nonzero result for the viscosity, but used a version of the model which had not yet been compared to flavor-dependent lattice QCD data. Importantly, fitting the data required us to introduce that $\tau$ dependence in the $w(\lambda,\tau_i)$ of~\eqref{eq:wtachyonform}, which is clearly stronger at small $\tau_i$ than that of the tachyon potential $\Vf(\lambda,\tau_i)$ of~\eqref{eq:Vfansatz}. We expect that this enhances the predictions for the bulk viscosity, making them better compatible with perturbative QCD results at high densities.

Finally, it will be interesting to study the physics of neutron stars and neutron star mergers by using the updated V-QCD model constructed here. We observed that turning on the strange quark mass enabled much smoother matching between the V-QCD quark matter EOS and nuclear matter EOSs than in earlier unflavored setups. This means that the nuclear to quark matter phase transition will become weaker, i.e., the latent heat will be significantly lower than in~\cite{Jokela:2018ers,Jokela:2020piw,Demircik:2021zll,Tootle:2022pvd,Ecker:2024kzs}. Earlier, the strong phase transition would destabilize all neutron stars having quark matter at their cores, but the updated EOS may lead to start with stable quark cores. This would lead to richer phenomenology for isolated neutron stars and neutron star mergers.

\acknowledgments

We thank for discussions T.~Demircik, C.~Ecker, Z.~Fang, S.~He, N.~Jokela, E.~Kiritsis, A.~Piispa, and E.~Pr\'eau.  M.~J. and T.~M. have been supported by an appointment to the JRG Program at the APCTP through the Science and Technology Promotion Fund and Lottery Fund of the Korean Government. M.~J. and T.~M. have also been supported by the Korean Local Governments -- Gyeong\-sang\-buk-do Province and Pohang City -- and by the National Research Foundation of Korea (NRF) funded by the Korean government (MSIT) (grant number 2021R1A2C1010834). T.M. has been supported by the DFG through the Emmy Noether Programme (project
number 496831614), through CRC 1225 ISOQUANT (project number 27381115).

\appendix

\section{Asymptotics near the boundary} \label{app:asymptotics}

The near-boundary asymptotics of the geometry depends on the weak-coupling expansion of the effective potential at zero tachyon and zero density,
\be
 V_g(\lambda) - x \Vf(\lambda,0) = \frac{12}{\ell^2}\left[1+v_1 \lambda + v_2 \lambda^2 + \mathcal{O}\left(\lambda^3\right)\right] \ ,
\ee
where $\ell$ is the AdS radius.
In terms of $v_1$ and $v_2$, we find for the scale factor and the dilaton the following expansions:
\begin{align}
 A(r)&=-\log \frac{r}{\ell}+\frac{4}{9 \log (r \Lambda)} +\\ & \quad+\frac{\left(\frac{95}{162}-\frac{32 v_2}{81 v_1^2}\right)+\left(-\frac{23}{81}+\frac{64 v_2}{81 v_1^2}\right) \log (-\log (r \Lambda))}{(\log (r \Lambda))^2}+\mathcal{O}\left(\frac{1}{(\log (r \Lambda))^3}\right)\ , \nonumber \\
v_1 \lambda(r)&=-\frac{8}{9 \log (r \Lambda)}+\frac{\left(\frac{46}{81}-\frac{128 v_2}{81 v_1^2}\right) \log (-\log (r \Lambda))}{(\log (r \Lambda))^2}+\mathcal{O}\left(\frac{1}{(\log (r \Lambda))^3}\right)\ . &
\label{eq:lambdaUV}
\end{align}
which also defines the energy scale $\Lambda$. 

The asymptotics of the tachyon depends on the expansions of $\kappa$ as well as the mass of the tachyon, obtained from expanding the potential $\Vf(\lambda,\tau_i)$ to second order in the tachyon. However, we choose the potentials such that this expansion is simply 
\begin{equation}
\Vf(\lambda,\tau_i) = V_{\mt{f}\lambda}(\lambda)\left[1-\tau_i^2 + \mathcal{O}\left(\tau_i^4\right)\right] \ .    
\end{equation}
In this case, we can define the expansion of $\kappa$ as 
\begin{equation}
    \kappa (\lambda) = \kappa_0 \left[1 + \kappa_1 \lambda + \mathcal{O}\left(\lambda^2\right)\right] \ .
\end{equation}
Consequently, the near-boundary expansion for the tachyon reads
\begin{equation} \label{eq:mqdef}
    \tau_i(r) = m_{i} r \ell \left[-\log(r\Lambda)\right]^{\gamma}\left[1+\mathcal{O}\left(\frac{1}{\log (r\Lambda)}\right)\right] \ ,
\end{equation}
where $\gamma = \frac{4}{3} + \frac{4\kappa_1}{3v_1}$. The factor of $\ell$ follows the normalization conventions from~\cite{Jarvinen:2011qe,Alho:2012mh}.

\section{Full background equations of motion} \label{app:EOM}

We complete here the discussion of Sec.~\ref{sec:thermo} by writing down
the full equations of motion. They are given by,
 \begin{align}
         \ddot{f} + 3 \dot{A} \dot{f} - \frac{ e^{2 A}}{\Nc} \sum_{i=1}^{\Nf} \frac{\sqrt{1+ e^{-2 A} f \dot{\tau}_i^2 \kappa}}{\sqrt{1 + K_i}} K_i V_{\mt{f}i} = 0
          \end{align}
          \begin{align}
    & 3 \ddot{A} + 3 \dot{A}^2 + 3 \frac{\dot{A} \dot{f}}{2 f} + \frac{2 \dot{\lambda}^2}{3 \dot{\lambda}^2}- \frac{e^{2 A} \Vg}{2 f} + \frac{ e^{2 A}}{2 \Nc f} \sum_{i=1}^{\Nf} V_{\mt{f}i} \sqrt{1+ e^{-2 A} f \dot{\tau}_i^2 \kappa} \sqrt{1 + K_i} = 0 
    \end{align}
    \begin{align}
    & \frac{\ddot{\lambda}}{\lambda} - \frac{\dot{\lambda}^2}{\lambda^2} + \frac{\dot{f} \dot{\lambda}}{f \lambda} + \frac{3 \dot{A} \dot{\lambda}}{\lambda} +\frac{3 e^{2 A}}{8 f} \lambda \partial_{\lambda} \Vg  & \\ 
    &- \frac{3}{8\Nc}\sum_{i=1}^{\Nf} \frac{ \left(V_{\mt{f}i}\sqrt{1+ e^{-2 A} f \dot{\tau}_i^2 \kappa} \right) \lambda  e^{2A}  }{ f \sqrt{1+ K_i}} \left(\partial_\lambda \ln \left( V_{\mt{f}i} \sqrt{1+ e^{-2 A} f \dot{\tau}_i^2 \kappa}  \right) -  K_i \partial_\lambda \ln w_i \right)  \nonumber   = 0 
    \end{align}
    \begin{align}
    & (1+K_i) \ddot{\tau_i} + \left( \dot{\tau}_i^2 + \frac{e^{2 A}}{f \kappa}\right) \left( K_i \partial_{\tau_i} \ln w_i - \partial_{\tau_i} \ln V_{\mt{f}i} \right) + \dot{\tau}_i^3 \dot{A} e^{-2 A} f \kappa(1+K_i) \nonumber \\ & +  \left( \dot{\tau}_i \dot{\lambda} \left( \partial_\lambda \log V_{\mt{f}i} - K_i \partial_\lambda \log w_i \right) + 3 \dot{\tau}_i \dot{A} \right) \left( 1+ \dot{\tau}_i^2 e^{-2 A} f \kappa \right) \nonumber \\ & + (1+ K_i) \dot{\tau}_i \left(\dot{\kappa} + \dot{f} \right)\left( 1+ \frac{1}{2}\dot{\tau}_i^2 e^{-2 A} f \kappa \right) = 0
\end{align}
\begin{align} \label{eq:constraint}
    \frac{\dot{\lambda}^{2}}{\lambda^2} + 9 \dot{A}^2 + \frac{9}{4} \frac{\dot{A} \dot{f}}{f} - \frac{3 e^{2 A}}{4 f}\left( \Vg - \frac{1}{\Nc} \sum_{i} V_{\mt{f}i} \frac{\sqrt{1+K_i}}{\sqrt{1+ e^{-2 A} f \tau_i'^2 \kappa}}\right) = 0 
    \end{align}
where
\begin{equation}
    V_{\mt{f}i} \equiv \Vf(\lambda,\tau_i) \ , \qquad w_i \equiv w(\lambda,\tau_i) \ .
\end{equation}

\subsection{Symmetries of the equations of motion}

The equations of motion are invariant under various symmetry transformations that arise from symmetries of the geometry and the action:
\begin{enumerate}
    \item \textbf{Shift of the coordinate $r$.} Since none of the coordinates appears explicitly in the action, it is trivially invariant in arbitrary constant shifts of the coordinates, $x^M \mapsto x^M + x_0^M$. Consequently, the background equations of motion are invariant under the shifts of the holographic coordinate, $r \mapsto r + r_0$. We fix this freedom by requiring that the UV boundary of the geometry is located at $r=0$.
    \item \textbf{Scaling of the coordinates and units.} The line element~\eqref{metric} is invariant under the scaling 
    \begin{equation}
        x^M \mapsto \widetilde \Lambda x^M \ , \qquad A \mapsto A - \log \widetilde \Lambda \ ,
    \end{equation}
    which corresponds to overall change in the units of energy in field theory. Due to covariance, this transformation also leaves the action unchanged, and is therefore reflected as the invariance of the background equations of motion under the scalings $r \mapsto \widetilde \Lambda r$ and $A \mapsto A - \log \widetilde \Lambda$. We fix this symmetry by specifying the value of $\Lambda$ in the near-boundary expansion~\eqref{eq:lambdaUV}.
    \item \textbf{Scaling of $f$.} The line element~\eqref{metric} is also unchanged under another transformation involving the blackening factor $f$. This transformation can be chosen to be such that it leaves $r$ invariant:
    \begin{equation}
        t \mapsto \frac{t}{f_0} \ , \qquad x_i \mapsto \frac{x_i}{\sqrt{f_0}} \ , \qquad f \mapsto f_0 f \ , \qquad A \mapsto A + \frac{1}{2}\log f_0 \ . 
    \end{equation}
    This is reflected in the invariance of the background equations of motion under $f \mapsto f_0 f$ and $A \mapsto A + \frac{1}{2}\log f_0$. This symmetry is fixed by requiring that $f$ approaches one at the boundary, which essentially fixes the speed of light in the field theory to one.
\end{enumerate}

\section{Potential sets} 

In this Appendix we present the full formulas for the potentials we use in this article. We present the Ansatz, and the parameters given by the comparison to the lattice data.

The complete Ansatz is given by 
\begin{align}
& V_\mt{g}(\lambda)=12\,\biggl[1+V_{g,1} \l+{V_{g,2}\lambda^2
\over 1+c_\l \l/\l_0}+V_\mathrm{IR} e^{-\l_0/(c_\l\l)}\left(\frac{c_\l\l}{\l_0}\right)^{4/3}\!\!\sqrt{\log(1+c_\l\lambda/\l_0)}\biggr]\  \\
\label{Vfan}
& V_\mt{f}(\lambda,\tau_i) = V_{\mt{f}\lambda}(\l) \left(1+
\tau_i^4\right)^{\tau_p} \exp\left(-\tau_i^2\right)   \\
&V_{\mt{f}\lambda}(\l)= W_0 + W_1 \l +\frac{W_2 \l^2}{1+c_\lambda\l/\l_0} + 12 W_\mathrm{IR}
e^{-\l_0/(c_\lambda\l)}(c_\lambda\l/\l_0)^{2}\\
\label{aan}
&\kappa(\l) = \kappa_0 \biggl[1+ \kappa_1 \lambda+
\bar \kappa_0 \left(1+\frac{\bar \kappa_1 \l_0}{c_\lambda\l} \right) e^{-\l_0/(c_\lambda\l) }\frac{(c_\lambda\l/\l_0)^{4/3}}{\sqrt{\log(1+c_\lambda\lambda/\l_0)}}\biggr]^{-1}  \\
\label{wan} &w(\l,\tau_i) =  \frac{w_0}{1+ \beta_s \tanh({\gamma_s^2 \tau_i^2})}\biggl[1 + \frac{w_1 {c_\lambda}\l/\l_0}{1+ {c_\lambda}\l/\l_0}  + \bar w_0
e^{-\l_0/(c_w\l)}\frac{(c_w\l/\l_0)^{4/3}}{\log(1+c_w\lambda/\l_0)}\biggr]^{-1} \ .
\end{align}
The function $\Vg$ is not refitted here but we use the parameters from~\cite{Jokela:2018ers}:
\begin{align}
    V_{g,1} = \frac{11}{27 \pi^2}, ~~~~ V_{g,2} = \frac{4619}{46656 \pi^4}, ~~~~ \lambda_0 = 8 \pi^2, ~~~~ V_\mathrm{IR} = 2.05, ~~~~ c_{\lambda}=3.
\end{align}

\subsection{Potential parameters} \label{app:st_pots}

The parameters of the other functions $V_f$, $\kappa$ and $w$ have been refitted to get the new set of potentials; \textit{standard potential} (\SP) and \textit{alternative potential} (\AP) based on the earlier sets of potentials, i.e., potentials 9 and potentials 7 of~\cite{Jokela:2018ers}, respectively. The parameters are obtained from the fit to the thermodynamics at $\mu = 0$, and to the light and strange quark susceptibilities. The Tables~\ref{tab:Pota} and~\ref{tab:Potb} below shows the parameters for the two potentials.

\begin{table}[H]
\centering
\begin{tabular}{|c|c|c|c|c|c|c|c|c|}
\hline
Potentials &$W_0$ & 
$\Lambda_\mt{UV}/$MeV & $W_\mt{IR}$ & $\bar{\kappa}_0$ & $ \bar{\kappa}_1$
& $45 \pi^2 M^3 \ell^3 /\left( 1+7/4 \right)$  \\
\hline
\SP & 5.886 & 
158.155 & 1 & 3.35 & 0.2
& 1.22   \\
\hline
\AP & 2.5 & 
210.76 & 0.9 & 1.8 & 0.2 
& 1.32 \\
\hline
\end{tabular}
\caption{Fit to thermodynamics for $\mu = 0$. Here $\ell$ is the UV AdS radius and $M$ is the 5-dimensional Planck mass normalized to give the Stefan-Boltzmann law for the pressure at higher temperature. $\Lambda_\mt{UV}$ is the energy scale of the background solution.}
\label{tab:Pota}
\end{table}

\begin{table}[H]
\centering
\begin{tabular}{|c|c|c|c|c|c|c|c|c|c|}
\hline
Potentials & $w_0$ & $c_w$ & $\bar{w}_0$  & 
$w_1$ & 
$\beta_s$ & $ \gamma_s$ \\
\hline
\SP & 0.81 & 1.1 & 12 & 
0.4  
&0.65 & 10   \\
\hline
\AP & 0.99 & 1.32 & 7.4  & 0.9  
&0.68 & 6   \\
\hline
\end{tabular}
\caption{Potential parameters obtained by fit to the light and strange quark susceptibility for $\mu = 0$.}
\label{tab:Potb}
\end{table}

In addition, the parameter $\kappa_0$ and $\kappa_1$ is given by,
\begin{align}
    \kappa_0 = \frac{3}{2} - \frac{W_0}{8}, \,\,\,\, \kappa_1 = \frac{11}{24 \pi^2}
\end{align}
and the parameters $W_1$ and $W_2$ are given by,
\begin{align}
    W_1 = \frac{8+3 W_0}{ 9 \pi^2}, ~~~~ W_2 = \frac{6488+999 W_0}{1552 \pi^2}.
\end{align}
These parameters are not fitted to lattice data but their values are determined by comparing to the RG flow in QCD perturbation theory~\cite{Jarvinen:2011qe}.

\bibliographystyle{JHEP}
\bibliography{references}

\end{document}